\def\isarxiv{1}      

\newif\ifarxiv
\newif\ifnotarxiv

\ifdefined\isarxiv
\arxivtrue
\fi

\ifarxiv\notarxivfalse
\else\notarxivtrue
\fi

\ifarxiv
\documentclass[sigconf]{acmart}
\fi

\ifnotarxiv
\documentclass[letterpaper,twocolumn,10pt]{article}
\usepackage{usenix-2020-09}
\fi

\usepackage[textsize=footnotesize]{todonotes}
\presetkeys{todonotes}{fancyline}{}
\usepackage{xcolor}
\usepackage{caption}
\usepackage{subcaption}
\usepackage{listings}
\usepackage{cleveref}
\usepackage{paralist}
\usepackage{enumitem}
\usepackage{pifont}
\usepackage{xspace}
\usepackage{comment}
\usepackage{fancyvrb}
\usepackage{graphicx,txfonts}
\usepackage{multirow}
\usepackage{float}
\usepackage{tikz}
\usepackage{multicol}
\usepackage{verbatim}
\usepackage{framed}

\ifnotarxiv
\setenumerate{noitemsep}
\setlist{nosep}
\fi

\newif\ifabridged
\newif\ifnotabridged
\newif\ifanonymous
\newif\ifnotanonymous
\newif\ifreviewer
\newif\ifnotreviewer

\ifdefined\isabridged
\abridgedtrue
\fi

\ifabridged\notabridgedfalse
\else\notabridgedtrue
\fi

\ifdefined\isanonymous
\anonymoustrue
\fi

\ifanonymous\notanonymousfalse
\else\notanonymoustrue
\fi

\ifdefined\isreviewer
\reviewertrue
\fi

\ifreviewer\notreviewerfalse
\else\notreviewertrue
\fi

\newenvironment{changed}{}{}

\newcommand{\dOne}{\ding{182}\xspace}
\newcommand{\dTwo}{\ding{183}\xspace}
\newcommand{\dThree}{\ding{184}\xspace}
\newcommand{\dFour}{\ding{185}\xspace}
\newcommand{\dFive}{\ding{186}\xspace}
\newcommand{\dSix}{\ding{187}\xspace}
\newcommand{\dSeven}{\ding{188}\xspace}
\newcommand{\dEight}{\ding{189}\xspace}
\newcommand{\dNine}{\ding{190}\xspace}
\newcommand{\dCOne}{\ding{192}\xspace}
\newcommand{\dCTwo}{\ding{193}\xspace}
\newcommand{\dCThree}{\ding{194}\xspace}
\newcommand{\dCFour}{\ding{195}\xspace}
\newcommand{\dCFive}{\ding{196}\xspace}
\newcommand{\dCSix}{\ding{197}\xspace}
\newcommand{\dCSeven}{\ding{198}\xspace}
\newcommand{\dCEight}{\ding{199}\xspace}
\newcommand{\dCNine}{\ding{200}\xspace}

\newcommand{\unicONE}{\includegraphics[scale=0.33]{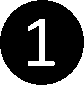}}
\newcommand{\unicTWO}{\includegraphics[scale=0.33]{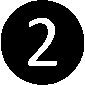}}
\newcommand{\unicTHREE}{\includegraphics[scale=0.33]{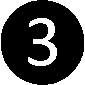}}
\newcommand{\unicFOUR}{\includegraphics[scale=0.33]{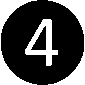}}
\newcommand{\unicFIVE}{\includegraphics[scale=0.33]{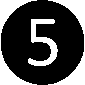}}
\newcommand{\unicSIX}{\includegraphics[scale=0.33]{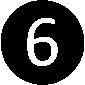}}
\newcommand{\unicSEVEN}{\includegraphics[scale=0.33]{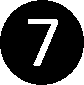}}
\newcommand{\unicEIGHT}{\includegraphics[scale=0.33]{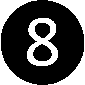}}
\newcommand{\unicNINE}{\includegraphics[scale=0.33]{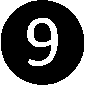}}
\newcommand{\unicTEN}{\includegraphics[scale=0.33]{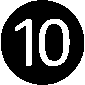}}
\newcommand{\unicELEVEN}{\includegraphics[scale=0.33]{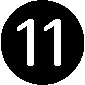}}
\newcommand{\unicTWELVE}{\includegraphics[scale=0.33]{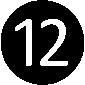}}

\newcommand{\unicFOURTEEN}{\includegraphics[scale=0.33]{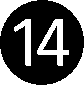}}
\newcommand{\unicFIFTEEN}{\includegraphics[scale=0.33]{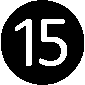}}

\newcommand{\SDRoB}{SDRoB\xspace}
\newcommand{\rootdomain}{root domain\xspace}
\ifnotarxiv
\renewcommand{\paragraph}{\noindent\textbf}
\fi
\newcommand{\isolated}{inaccessible\xspace}
\newcommand{\nonisolated}{accessible\xspace}
\newcommand{\specialcell}[2][c]{%
  \begin{tabular}[#1]{@{}c@{}}#2\end{tabular}}

\newcommand{\captionenumber}[1]{\begin{inparaenum}[1)] #1 \end{inparaenum}}
\newcommand{\captionenumalpha}[1]{\begin{inparaenum}[(a)] #1 \end{inparaenum}}

\definecolor{mGreen}{rgb}{0,0.6,0}
\definecolor{mGray}{rgb}{0.5,0.5,0.5}
\definecolor{mPurple}{rgb}{0.58,0,0.82}
\definecolor{backgroundColour}{rgb}{0.95,0.95,0.92}

\newcommand\LSTSize{\fontsize{7}{7.2}\selectfont}
\newcommand*\LSTfont{\LSTSize\ttfamily\SetTracking{encoding=*}{-0}\lsstyle}

\lstdefinestyle{CStyle}{
    backgroundcolor=\color{backgroundColour},
    commentstyle=\color{mGreen},
    keywordstyle=\color{magenta},
    numberstyle=\tiny\color{mGray},
    stringstyle=\color{mPurple},
    basicstyle=\LSTfont,
    breakatwhitespace=false,
    breaklines=true,
    captionpos=b,    
    escapeinside={\%*}{*)},
    keepspaces=true,
    numbers=left,
    numbersep=5pt,
    showspaces=false,
    showstringspaces=false,
    showtabs=false,
    tabsize=2,
    language=C,
    classoffset=1,
    morekeywords={enter_domain,sdrob_call,sdrob_enter,sdrob_malloc,sdrob_exit,sdrob_init,sdrob_destroy,sdrob_free,sdrob_deinit,sdrob_dprotect},
    keywordstyle=\color{orange},
    classoffset=2,
    morekeywords={EXECUTION_DOMAIN,ISOLATED_DOMAIN,RETURN_TO_CURRENT,DATA_DOMAIN,SUCCESSFUL_RETURNED,EXECUTION_DOMAIN,NONISOLATED,RETURN_HERE,RETURN_TO_PARENT,NO_HEAP_MERGE,HEAP_MERGE,OK,OK,MALLOC_FAILED,WRITE_ENABLE,READ_ENABLE,ACCESSIBLE,INACCESSIBLE,ACCESSIBLE_DOMAIN,INACCESSIBLE_DOMAIN,udi_t},
    keywordstyle=\color{brown},
    classoffset=0,
}

\definecolor{terminalbgcolor}{HTML}{330033}
\definecolor{terminalrulecolor}{HTML}{000099}
\lstdefinestyle{ConsoleStyle}{
    backgroundcolor=\color{terminalbgcolor},
    basicstyle=\color{white}\fontfamily{fvm}\footnotesize\selectfont,
    breakatwhitespace=false,
    breaklines=true,
    captionpos=b,
    commentstyle=\color{mGreen},
    deletekeywords={...},
    escapeinside={\%*}{*)},
    extendedchars=true,
    frame=single,
    keepspaces=true,
    keywordstyle=\color{blue},
    morekeywords={*,...},
    numbers=none,
    numbersep=5pt,
    framerule=2pt,
    numberstyle=\color{mGray}\tiny\selectfont,
    rulecolor=\color{terminalrulecolor},
    showspaces=false,
    showstringspaces=false,
    showtabs=false,
    stepnumber=2,
    stringstyle=\color{mPurple},
    tabsize=2
}

\usetikzlibrary{decorations.pathreplacing,positioning,arrows,tikzmark}

\tikzset{
  comment/.style={
    draw=none,
    text=black,
    align=left,
    inner sep=2,
    outer sep=1,
    font=\rmfamily
  },
  none/.style={
    draw=none,
    text=black,
    align=left,
    inner sep=2,
    outer sep=1,
    font=\rmfamily
  },
}

\ifarxiv
\setcopyright{none} 
\acmConference[Anonymous Submission to ACM CCS 2022]{ACM Conference on Computer and Communications Security}{Due 2 May 2022}{Los Angeles, U.S.A.}
\acmYear{2022}
\fi

\begin{document}

\ifarxiv
\title{Unlimited Lives: Secure In-Process Rollback with Isolated Domains}
\ifnotanonymous
\author{Merve G\"{u}lmez}		
\affiliation{
  \institution{Ericsson Security Research}
  \country{Kista, Sweden}
}
\affiliation{
    \institution{imec-Distrinet, KU Leuven}
    \country{Leuven, Belgium}
}
\email{merve.gulmez   
@kuleuven.be}

\author{Thomas Nyman}
\affiliation{
  \institution{Ericsson Product Security}
  \country{Jorvas, Finland}
}
\email{thomas.nyman
@ericsson.com} 

\author{Christoph Baumann}
\affiliation{
    \institution{Ericsson Security Research}
    \country{Kista, Sweden}
}
\email{christoph.baumann
@ericsson.com} 

\author{Jan Tobias M\"{u}hlberg}
\affiliation{
    \institution{imec-Distrinet, KU Leuven} 
    \country{Leuven, Belgium}}
\affiliation{    
    \institution{Université Libre de Bruxelles} 
    \country{Brussels, Belgium}}
\email{jan.tobias.muehlberg@ulb.be} 
\fi
\fi

\ifnotarxiv
\title{\Large \bf Unlimited Lives: Secure In-Process Rollback with Isolated
Domains}

\ifnotanonymous
\author{
{\rm Merve Turhan}\\
Ericsson Security Research\\
imec-DistriNet, KU Leuven\\
\and
{\rm Thomas Nyman}\\
Ericsson Product Security\\
\and
{\rm Christoph Baumann}\\
Ericsson Security Research
\and
{\rm Jan Tobias M\"{u}hlberg}\\
imec-DistriNet, KU Leuven
} 
\fi
\fi

\ifnotarxiv
\maketitle
\fi

\begin{abstract}
The use of unsafe programming languages still remains one of the
major root causes of software vulnerabilities. Although well-known defenses that
detect and mitigate memory-safety related issues exist, they don't address the
challenge of software \emph{resilience}, i.e., whether a system under attack
can continue to carry out its function when subjected to malicious input.
We propose \emph{secure rollback of isolated domains} as an efficient
and secure method of improving the resilience of software targeted by run-time attacks.
We show the practicability of our methodology by realizing a software library for
Secure Domain Rollback (\SDRoB) and demonstrate how \SDRoB can be applied to real-world software.
\end{abstract}

\ifarxiv 
\def\UrlFont{\sffamily\footnotesize}
\maketitle
\def\UrlFont{\ttfamily\footnotesize}
\fi

\section{Introduction}\label{sec:introduction}
Software written in unsafe programming languages can suffer from various
memory-related vulnerabilities~\cite{erlingsson2010low} that allow run-time
attacks, such as control-flow attacks and non-control-data
attacks~\cite{nondatacontrolattack}, to compromise program behavior. Attackers
use such run-time attacks to gain access to vulnerable software and
systems. According to the Google \ifnotabridged Project Zero \fi  "0day In the
Wild" dataset over 70\% of the zero-day vulnerabilities between July 2014
and \begin{changed}June\end{changed} 2022 can be attributed to memory-safety
issues~\cite{project0day}.  Research into run-time attacks has, during the past
30 years, led to an ongoing arms race between increasingly sophisticated attacks
and run-time defenses to mitigate such attacks~\cite{Szekeres13}.  Today, major
operating systems (OSs) provide such mitigations by default.  This
includes non-executable stack and heap areas, address-space-layout randomization
(ASLR)~\cite{ASLR}, toolchain hardening options such as stack
canaries~\cite{stackcanaries}, and hardware-enforced control-flow integrity
(CFI)~\cite{declercq2017survey}.  However, virtually all currently known
defenses mitigate detected attacks by terminating the victim
application\ifabridged~\cite{Abadi09,Burow17,Castro09,Cowan98,Erlingsson06,Larsen14,Liljestrand19a,Liljestrand21,Mao11,Niu14,Niu15,Park19,Schwartz11,Serebryany19,Szekeres13,vanderVeen15,Wahbe93}\else~\cite{Abadi09,Bhatkar08,Burow17,Burow19,Brown17,Cadar08,Castro06,Castro09,Cheng17,Cheng19,Chiueh01,Crane15,Cowan98,Dang15,Devietti08,Ding17,Duck16,Duck17,Erlingsson06,Feng19,Giffin02,Giffin04,Habibi15,Hu18,Intel-CET,Kc03,Kuznetsov14,Kwon13,Larsen14,Liljestrand19a,Liljestrand19e,Liljestrand21,Mao11,Mashtizadeh15,Nagarakatte09,Nebenzahl06,Niu14,Niu15,Park19,Prasad03,Schlesinger11,Schwartz11,Serebryany12,Serebryany19,Seshadri05,Song16,Szekeres13,vanderVeen15,Wahbe93,Watson15,Woodruff14,Tice14,Tsampas17,Xu02,Zhang15}\fi.
Thus, even though applications are hardened against run-time attacks, the
response can still be leveraged by attackers to create temporary
denial-of-service conditions while the application is restarted, or to bypass
security controls by resetting volatile system state, e.g., counts of failed
login attempts.

Service-oriented applications are at particular risk as even a temporary failure
in a critical component can affect a large number of clients.  An example for
such an application is \emph{Memcached}, a general-purpose distributed
memory-caching system, which is commonly used to speed up database-driven
applications by caching database content.

\paragraph{Contributions.}
To address the limitation of current defenses and improve the resilience of
software that is being targeted by run-time attacks we propose \emph{secure
rollback of isolated domains}. Secure rollback allows the state of a victim
application under attack to be restored to a prior state, known to be unaffected
by an ongoing run-time attacks. This is possible by leveraging hardware-assisted
software fault isolation (SFI) to compartmentalize the application into
distinct \emph{domains} that limit the effects of run-time attacks to isolated
memory compartments. An application can be instrumented to isolate, e.g.,
"high-risk" code that operates with untrusted input in a secure in-process
sandbox and roll back the application state if an attack is detected against
sandboxed code. Domains can be nested to allow for efficient and
secure rollback in different software architectures and use cases.

We show the practicability of our methodology by realizing a software
library for \emph{secure domain rollback} (\SDRoB) for commodity 64-bit x86
processors with \emph{protection keys for userspace} (PKU)~\cite{Intel64,AMD-AMD64} and demonstrate
how \SDRoB can be applied to real-world software in \begin{changed}case studies on Memcached,
a popular distributed memory-cache system (\Cref{sec:memcached}), the NGINX web server (\Cref{sec:nginx}), and OpenSSL (\Cref{sec:ssl}).\end{changed}
In summary, the contributions of this paper are:
\begin{itemize}[leftmargin=*]
  \item \emph{Secure Rollback of Isolated Domains} is a novel
scheme to improve software resilience against run-time attacks by
rolling back the state of a victim application (\Cref{sec:system_design}).
  \item We explore different \emph{design patterns for compartmentalization and
    rollback} and discuss their applicability to retrofit existing software with secure rollback (\Cref{sec:domainlifecycle}-\ref{sec:multithreading}).
  \item We provide \SDRoB, a realization of secure rollback  for commodity
64-bit x86 processors with PKU (\Cref{sec:implementation}).
  \item %
\begin{changed}We apply \SDRoB in three case studies (Memcached, NGINX, and OpenSSL, \Cref{sec:performance}) and show that it can be used with minor changes
    to application code, exhibiting in benchmarks on Memcached and NGINX a worst
case performance overhead <7.2\%, negligible overhead (2\%--4\%) in realistic
multi-processing scenarios, and negligible memory overhead
(0.4\%--3\%) \end{changed}. We assess the security and applicability of our
approach in \Cref{sec:security_ev} and \Cref{sec:applicability}.
\end{itemize}


\section{Background}\label{sec:background}
Memory-safety issues cause vulnerabilities such as buffer
overflows, use-after-free, and format string
vulnerabilities~\cite{Szekeres13}.  Today, countermeasures such as
W$\oplus$X~\cite{Schwartz11}, ASLR~\cite{Larsen14}, stack
canaries~\cite{Cowan98} and CFI~\cite{Abadi09} that mitigate memory
vulnerabilities are widely deployed by all major OSs.

However, the current state-of-practice in run-time defenses focuses on
detecting attacks and terminating the offending processes. This effectively
prevents attackers from leveraging memory vulnerabilities as stepping
stones for privilege escalation, remote code execution, or data exfiltration,
but disregards availability concerns in favor of disrupting the
attack's kill chain. This is acceptable for protecting end users from threats,
e.g., restarting a browser is a minor inconvenience.
However, in high-availability applications any measure which may
cause service disruption necessitates overall system resilience to be provided
through redundancy and load balancing.

\ifnotabridged
It may also be desirable to collect diagnostics data from the application when
an attack is detected to aid in root-cause analysis. Instant termination of an
application hinders the collection of diagnostics data, except perhaps a core
dump, but collecting data from a process which is under the attacker's control
poses yet another challenge.
\fi

\begin{changed}
\subsection{Software Fault Isolation}\label{sec:sfi}

Software fault isolation (SFI)~\cite{Tan17} is a technique for
establishing logical protection domains within a process through program
transformations. SFI instruments the program to intermediate memory accesses to
ensure they do not violate domain boundaries. Since transitioning from one
domain to another stays within the same process SFI solutions can offer better
run-time efficiency compared to traditional process isolation, especially in use
cases where domain transitions are frequent. SFI has been successfully deployed
for sandboxing plug-ins in the Chrome browser~\cite{Yee09}, isolating OS kernel~\cite{roessler2021} and
modules~\cite{Erlingsson06, Castro09, Mao11}, as well as code accessed through
foreign function interfaces in managed language runtimes~\cite{Sun12}. 

SFI enforcement can be realized in different ways. The principal method to
realize SFI for native code binaries is the use of an \emph{inline reference
monitor} through binary-~\cite{Erlingsson06} or compiler-based
rewriting~\cite{Wahbe93, Castro09, Mao11, PtrSplit} of the application binary.
Recent SFI approaches leverage hardware-assistance~(\Cref{sec:mpk}) to further
improve enforcement efficiency~\cite{Rivera16,Koning17,Hedayati19,Vahldiek-Oberwagner19,Melara19,Sung20,Lefeuvre21,Schrammel20,Wang20,Voulimeneas22,Kirth22,Jin22,Chen22}. Existing approaches to SFI share the drawback of memory vulnerability
countermeasures as they respond to detected domain violations by terminating the
offending process.
\end{changed}

\ifnotabridged
\subsection{Checkpoint \& Restore}\label{sec:check_restore}

Application checkpoint \& restore~\cite{restoring} is a technique for increasing
system resilience against failure. It involves saving the state of a running
process periodically or before a critical operation, so that a failed process
can later be restarted from the checkpoint. The cost of this depends on the
amount of data that is needed to capture the system's state and the
checkpointing interval.

Several studies have focused on optimizing
checkpointing~\cite{optimizechekpoint, Young1974AFO, checkpointrestart}.
Checkpoints can be created at system level or application level. At system
level, checkpointing needs to capture a complete reproduction of the
application's memory as well as other attributes, such as sockets, open files,
and pipes. At application level, checkpointing requires additional functionality
inserted into the application itself to facilitate checkpoint \& restore.

Secure checkpointing schemes~\cite{seccheck} generally leverage cryptography to
protect the integrity and confidentiality of checkpoint data at rest. However
they generally do not consider attacks that may tamper with checkpointing
code at the application level. Furthermore, the cost of bulk encryption of
checkpoint data is too high for latency-sensitive applications, e.g., network traffic
processing or distributed caches unless the system provides load balancing and
redundancy.

In this work, we avoid the pitfalls of checkpoints that reproduce process memory
by leveraging hardware-assisted fault isolation to partition an application
process intro distinct, isolated domains. This compartmentalization facilitates
secure rollback of application state by isolating the effects of memory errors.
This enables rollback to application states that precede the point of failure in
the application's call graph and are unaffected by a caught and contained error.
\fi

\subsection{Memory Protection Keys}\label{sec:mpk}

Memory protection keys (MPK) provide an access control mechanism that augments
page-based memory permissions. MPK allows memory access permissions to be
controlled without the overhead of kernel-level modification of page table
entries (PTEs). On 64-bit x86 processors, protection keys for userspace code
(PKU) are supported 
\ifabridged in Intel's~\cite{Intel64} and AMD's~\cite{AMD-AMD64}
microarchitectures\else since Intel's Skylake~\cite{Intel64} and AMD's Zen
3~\cite{AMD-AMD64} microarchitectures\fi. 
 \ifnotabridged Future Intel processors will add support for memory
protection keys for supervisor mode (PKS)~\cite{Intel64}. \fi  
Similar hardware mechanisms are also available in ARMv8-A~\cite{ARMv8-A}, IBM
Power~\cite{IBM-StorageProtectKeys}, HP PA-RISC~\cite{HP-PA-RISC} and
Itanium~\cite{Intel-IA64} processor architectures.

\ifnotabridged
\Cref{fig:PKU} illustrates the component of PKU on 64-bit x86 processors.
Each memory page is associated with a 4-bit protection key stored in the page's
PTE \dOne. The access rights to memory associated with each
protection key are kept in a \emph{protection key rights register} (PKRU)~\dTwo.
The PKRU allows write-disable (WD) and access-disable (AD) policies to be
configured for protection keys. These policies are enforced by hardware on
each memory access.
\fi

\ifabridged
Each memory page is associated with a 4-bit protection key stored in the page's
PTE on on 64-bit x86 processors. The access rights to memory associated with each
protection key are kept in a \emph{protection key rights register} (PKRU).
The PKRU allows write-disable (WD) and access-disable (AD) policies to be
configured for protection keys. These policies are enforced by hardware on
each memory access.
\fi

Unlike MPK mechanisms for architectures such as ARMv8-A and IBM Power, which
limit access to the PKRU to privileged code, the PKRU
in 64-bit x86 is configurable from userspace. This allows domain transitions to
occur efficiently without involving the OS kernel. However, it also means that
PKU on its own cannot effectively enforce secure in-process isolation, but has
to be combined with mechanisms such as W$\oplus$X and CFI that can limit PKRU
access to code which is trusted to manage isolation. Existing work has shown
that compiler-based code rewriting~\cite{Koning17} or binary
inspection~\cite{Vahldiek-Oberwagner19} when combined with system call
filtering~\cite{Voulimeneas22}, or non-invasive hardware
extensions~\cite{Schrammel20}, provide sufficient security for PKRU access to
preclude bypassing PKU policies.

\ifnotabridged
\begin{figure}[tb]
    \centering
    \includegraphics[scale=0.35]{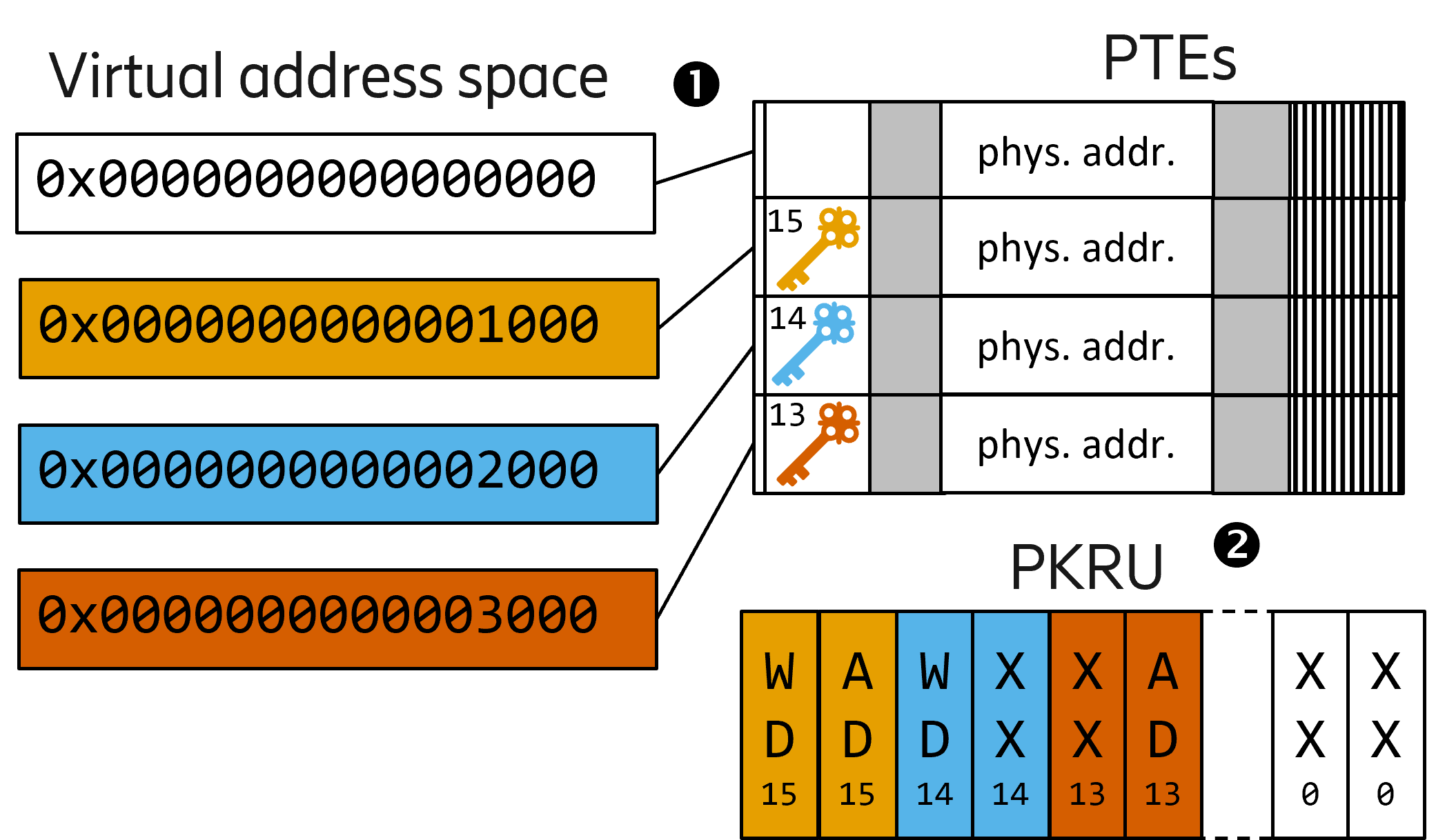}
    \caption{Overview of 64-bit x86 PKU components.}
    \label{fig:PKU}
\end{figure}
\fi

\section{Secure Domain Rollback}\label{sec:system_design}
We propose \emph{secure rollback of isolated domains}, a novel approach for
improving the resilience of userspace software against run-time attacks that
augments existing, widely deployed run-time defenses. First, we present our
threat model and system requirements (\Cref{sec:threatmodel}). Then we introduce
a motivating example (\Cref{sec:vulnprogram}) and use it to explain the
high-level idea behind the solution (\Cref{sec:highlevelidea}).
\Cref{sec:domainlifecycle,sec:domainnesting,sec:multithreading} delve deeper
into specific aspects of the design.

\subsection{Threat Model and Requirements}\label{sec:threatmodel}

\paragraph{Assumptions.} %
In this work, we assume that the attacker has arbitrary access to process
memory, but is restricted by the following assumptions about the system:

\begin{enumerate}[label=\textbf{A{\arabic*}}, leftmargin=*]
  \item\label{as:wxorx} A W$\oplus$X policy restricts the adversary from
    modifying code pages and performing code-injection.

  \item\label{as:hardening} The application is hardened against run-time attacks
    and can detect an attack in progress, but not necessarily prevent
    the attacker from corrupting process memory. 
    The general Linux protections (cf.~\Cref{sec:background}) are in place.
\end{enumerate}

We limit the scope of \ref{as:hardening} to software-level attacks.
Transient execution~\cite{Xiong21} and hardware-level attacks, e.g., fault
injection~\cite{Shepherd21}, rowhammer~\cite{Mutlu20} etc., which are generally
mitigated at hardware, firmware, or kernel level, are out of scope.

\newcommand{\reqRecovery}[0]{The mechanism must allow the application to
continue operation after system defenses (\ref{as:hardening}) detect an attack.}

\newcommand{\reqMemoryIntegrityA}[0]{The mechanism must ensure that the
integrity of memory after recovering from the detected attack is maintained.}

\newcommand{\reqMemoryIntegrityB}[0]{To facilitate \ref{req:recovery} and
\ref{req:memory_integrity} the application is compartmentalized into isolated
domains with the following requirements:}

\newcommand{\reqMemoryIntegrityC}[0]{Run-time attacks that affect one domain
must not affect the integrity of memory in other domains}

\newcommand{\reqIsolationMaintained}[0]{The attacker must not be able to
tamper with components responsible for
isolation or transitions between domains, or data
used as part of the rollback process.}

\paragraph{Requirements.} %
Our goal is to improve the resilience of an application against active
run-time attack that may compromise the integrity of the application's
memory. We introduce a mechanism for \emph{secure rollback} with 
requirements as follows:
\begin{enumerate}[label=\textbf{R{\arabic*}}, leftmargin=*]
  \item\label{req:recovery} \reqRecovery{}

  \item\label{req:memory_integrity} \reqMemoryIntegrityA{}
\end{enumerate}

\reqMemoryIntegrityB{}

\begin{enumerate}[label=\textbf{R{\arabic*}}, leftmargin=*,resume]
  \item\label{req:domain_isolation} \reqMemoryIntegrityC{}

  \item\label{req:isolation-maintained} \reqIsolationMaintained{}
\end{enumerate}

\subsection{A Vulnerable C Program}\label{sec:vulnprogram}

Latent memory vulnerabilities may exist undiscovered within applications until
they are set off by input that triggers the software defect.
When a defect is triggered, it is highly likely that it
causes the application's memory to become corrupt in unexpected ways.  As a
motivating example, consider the program in \Cref{lst:example_overflow} that
contains a "classic" buffer overflow vulnerability.  The memory vulnerability~\dCOne
is triggered by input that exceeds the size of the input buffer
\texttt{buf}.  As the input overflows the buffer it will corrupt the surrounding
stack frame, and eventually overflow into the main function's stack frame
where it will corrupt the local variable \texttt{sum}.

Modern C compilers guard against this particular vulnerability
\ifabridged
by emitting stack canaries that detect if a buffer overflow corrupts the stack.
\else
in two ways:
\begin{inparaenum}[1)]
\item by deprecating the \texttt{gets()} function in favor of \texttt{fgets()}
  with explicit array bounds checks,
\item by emitting stack canaries at stack frame boundaries which indicate if a
  buffer overflow corrupts a function's stack frame.
\end{inparaenum}
\Cref{lst:stack_smashing} shows how the hardened application is terminated at
run-time when the overflow is detected.
\fi
However, by the time the overflow is
detected the application's memory has already been corrupted, rendering the
application process unrecoverable.

\subsection{High-level Idea}\label{sec:highlevelidea}
The objective of the secure rollback mechanism is to recover the application's
execution state, after a memory defect has triggered, to a prior state before the
application's memory has been corrupted (\ref{req:memory_integrity}). Thus, the
application can resume its execution and continue to provide its services
without interruption (\ref{req:recovery}). To facilitate this, the application is compartmentalized
into separate, isolated domains that each execute in distinct memory
compartments allocated from the process's memory space.  Should the execution of
code inside a domain fail due to a memory defect, the memory that belongs to the
domain may be corrupt.  However, since the effects of the memory defect are
isolated to the memory belonging to the failing domain
(\ref{req:domain_isolation}, \ref{req:isolation-maintained}) the application's
execution can now be recovered by:
\begin{inparaenum}[1)]
\item discarding any affected memory compartments, 
\item unwinding the application's stack to a state prior before the offending
  domain began its execution, and
\item performing an application-specific error handling procedure that avoids
  triggering the same defect again, e.g., by discarding the potentially
  malicious input that caused it.
\end{inparaenum}

\Cref{lst:example_simple_domain} shows how the program from
\Cref{lst:example_overflow} is modified to benefit from secure rollback.  The
call to \texttt{get\_number()} has been wrapped by a call to the
\texttt{enter\_domain()} function~\dOne.  This wrapper function
is part of secure rollback instrumentation, here shown with simplified
arguments. It performs the following actions:
\begin{itemize}[leftmargin=*]
  \item Save information about the calling environment such as register
    values, including the stack and instruction pointers, and signal mask (in a
    manner similar to C \texttt{setjmp()}~\cite{setjmp})
    for later use by the rollback mechanism
  \item Reserve a portion of application memory to store the new domain's stack
    and heap, persisting for the duration the domain remains active
    (see \Cref{sec:domainlifecycle}).
  \item \begin{changed}Update the hardware-enforced memory access policy, granting access to
    the new domain's memory areas and protecting all other memory. Global
    data is set to read-only.\end{changed}
  \item Finally, invoke the function inside the new domain.
\end{itemize}

\begin{lstlisting}[float=t, style=CStyle, label={lst:example_overflow}, caption={A vulnerable C application.}]
int get_number() {
  char buf[BUFSIZE]; // BUFSIZE = 8
  gets(buf); %*\dCOne\hspace{-.8em}*)       // read input to buf
  return atoi(buf);  // return number as integer, or zero
}                    // if conversion is not possible

void main(void) {
	int sum = 0;       // sum of all inputs
	for (;;) {
		sum += get_number();
		printf("The sum so far: %d\n", sum);
	}
}
\end{lstlisting}

\ifnotabridged
\begin{lstlisting}[float, style=ConsoleStyle, label={lst:stack_smashing}, caption={A stack overflow is detected at run-time.}]
./a.out
AAAAAAAAA
*** stack smashing detected ***: terminated
Aborted (core dumped)
\end{lstlisting}
\fi

\begin{lstlisting}[float=t, style=CStyle, label={lst:example_simple_domain}, caption={The loop from \Cref{lst:example_overflow} equipped for rollback.}]
int ret;  // to store return value on normal domain exit
for(;;) {
  // run get_number in isolated domain
  if(enter_domain(&get_number, &ret) == OK) { %*\dOne*)
    sum += ret;  // normal domain exit   %*\dTwo*)
    printf("The sum so far: %d\n", sum);
  } else {  // abnormal domain exit %*\dThree*)
    printf("ERROR! Bad Input");
  }
}
\end{lstlisting}

The newly spawned domain can end its execution in one of two ways:
\begin{inparaenum}[1)]
  \item \emph{normal domain exit}, or
  \item \emph{abnormal domain exit}
\end{inparaenum}
A normal domain exit~\dTwo occurs when the application's execution flow
returns naturally from the code invoked inside the domain to the call site. This
means that the isolated code has completed successfully and application
execution is resumed outside the domain.  An abnormal domain exit occurs~\dThree
if the isolated code tries to access memory past the
confines of the domain's memory area, or a possible run-time attack is
detected. In an abnormal domain exit, the execution of the domain is halted, the
domain's memory is discarded, and the application execution is resumed by
restoring the calling environment from the information stored prior to invoking
the offending domain, effectively rolling back application state to the point
before the domain started executing.  In \Cref{lst:example_simple_domain}, the
effects of a buffer overflow inside \texttt{get\_number()} are limited to the
newly created domain; the main function's stack is unaffected by the
overflow, and the rollback mechanism can thus transfer control back to \texttt{main()}.

The caller learns a domain's exit status 
from \texttt{enter\_domain()}'s return value.  On abnormal domain exit the application is
expected to take an alternate action to avoid the conditions that lead to the
previous abnormal exit before retrying the operation. For example, a
service-oriented application can close the connection to a potentially malicious
client. The rollback of application state is limited
to the state of the application's memory.
Operations that have side-effects on the application's environment, e.g.,
reading from a socket, are still visible to the application after rollback.

In the following we explain design patterns for secure
rollback that apply to different software architectures.

\subsection{Domain Life Cycle} \label{sec:domainlifecycle}
As part of process initialization, all application memory, including stack,
heap, and global data, are assigned to the \emph{\rootdomain},
which forms the initial isolated domain where an application executes.
 Application subroutines are compartmentalized into \emph{nested domains}
 to create multiple recovery points from which
 rollbacks can be performed at any point during 
 execution. As
 the names suggests, domains can be nested in the sense that several domains can
 be entered subsequently starting from the \rootdomain, each with a dedicated
 rollback procedure (cf.~\Cref{sec:domainnesting}). 
\begin{changed} While read-only access from any nested domain to data in the parent domain may be allowed, writable access is forbidden in order to contain any memory safety violations to the nested domain. By default, nested domains have read access to the \rootdomain to enable reading global
variables.\end{changed} We envision two flavors of isolation with rollback for
application subroutines:
\begin{itemize}[leftmargin=*]
  \item \textbf{Protecting the application from a subroutine:} Code that may
    have undetected memory vulnerabilities, e.g., third-party software
    libraries, can be executed in a nested domain by instrumenting calls to the
    functionality so that they execute in their own domain and may be rolled
    back in case memory safety violations are detected.
  \item \textbf{Protecting a subroutine from its caller:} parts of the
    application that operate on sensitive data such as
    cryptographic keys can be isolated from vulnerabilities in its callers,
    preventing the leak and loss of such data. For example, functions for
    encryption, decryption, or key derivation  from the OpenSSL library can be
    isolated in their own nested domain to protect the application's
    cryptographic keys if a fault occurs in a calling nested domain.
    \Cref{lst:openssl_wrapper} in \Cref{sec:api} \begin{changed}and the calling
    nested domain in \Cref{sec:openssl_example}\end{changed} show a concrete
    example of isolating OpenSSL.
\end{itemize}

When the application's execution flow enters any isolated subroutine, a nested
domain is created and assigned a substack and subheap area. To support the two
application scenarios described above, we identify two design patterns for
nested domains. The type of a domain determines how it continues its
life cycle, in particular what happens upon domain exit.
 
\paragraph{Persistent Domains}
A persistent domain retains all assigned memory areas even after the
application's execution flow returns from the persistent nested domain to the
parent domain. Another code path may enter the persistent domain again, at which
point access is granted to memory areas that belong to the persistent domain.
Modules that maintain state information across invocations should be isolated in
a persistent domain so that their state is not lost after a normal domain
exit. For instance, some software libraries may encapsulate such state by
creating a "context" object that decouples domain-specific data from business
logic. One strategy for compartmentalizing such libraries is to ensure each
distinct context is allocated in different persistent domains. A good example of
this pattern is OpenSSL which can be instantiated multiple times within an
application using different contexts. Assigning one persistent domain per
\begin{changed}concurrent\end{changed} context therefore ensures cryptographic
keys associated in memory with one context remain isolated from other domains.

On abnormal exits from any domain, the rollback mechanism is triggered and all
state of the domain is discarded. Note that for persistent domains this may have
serious repercussions for an application, if the program state depends on the
persistent state of the isolated library. When, for example, the abnormal exit
leads to the loss of session keys for a TLS connection, the application may need
to recover by re-initializing the affected context and close connections that
were handled in the lost context.

Depending on the application, the parent domain may or may not be given access
to a persistent nested domain's memory. For instance, in the case of
cryptographic libraries, access to the persistent domain's memory from any other
domain should be blocked to protect sensitive data stored by the library. In
such cases data cannot be directly passed between the caller and callee in
distinct domains and shared data, e.g., call arguments and results need to be
copied between the nested domain and its caller via a designated shared memory
area.  This is similar to how data is passed between different protection
domains in, e.g., Intel SGX
enclaves~\cite{Intel-SGX}.

\paragraph{Transient Domains} 
Memory areas assigned to transient nested domains persist until the
application's execution flow returns from such a domain to the parent domain at
which point the stack as well as \emph{unused} heap memory areas assigned to the
transient nested domain are discarded. For this the rollback mechanism needs to
keep track of memory allocations in a nested domain. Any allocated memory in a
transient nested domain's heap area can be merged back to the parent domain's
heap area or discarded upon a normal domain exit, depending on the specific
application scenario.
\ifabridged
\begin{changed}
Generally, transient domains are most useful for functionality that is called
only once during a typical program invocation.
\end{changed}
\fi
\ifnotabridged
In the example in~\Cref{lst:mergingarea}, execution enters a nested domain
\dOne where \texttt{get\_name()} \dTwo dynamically allocates a buffer in
heap memory \dThree and returns it as a result to \texttt{main()}. In a normal
domain exit \dFour, the active allocation of the transient domain is merged to
the \rootdomain where \texttt{main()} lives. This means that control of the
buffer allocated in the transient domain and pointed to by \texttt{ret} is
transferred. The developer can free this memory later \dFive. Unused
heap area of the nested domain and the stack are discarded automatically by the
secure rollback mechanism. On an abnormal domain exit \dSix substack area and
subheap area assigned to the transient nested domain are discarded by the
rollback mechanism already and no further clean-up by the developer is
necessary.

\begin{lstlisting}[style=CStyle, label={lst:mergingarea}, caption={Domain Merging Area Example}]
void get_name() { %*\dTwo*)
  void *buf;
  buf = malloc(BUFSIZE); %*\dThree*)// BUFSIZE = 8  
  gets(buf);            
  return buf; 
}                    
void main(void) { 
  void *ret; // holds the return value
  for(;;) {
    if(enter_domain(&get_name, ret) == OK){ %*\dOne*)
      printf("%s", ret); // normal domain exit %*\dFour*)
      free(ret); %*\dFive*)
    } else {  // abnormal domain exit %*\dSix*)
      printf("ERROR! Bad Input");     
  }
}  
\end{lstlisting}
\fi
\begin{figure*}[t]
  \centering \includegraphics[scale=0.42]{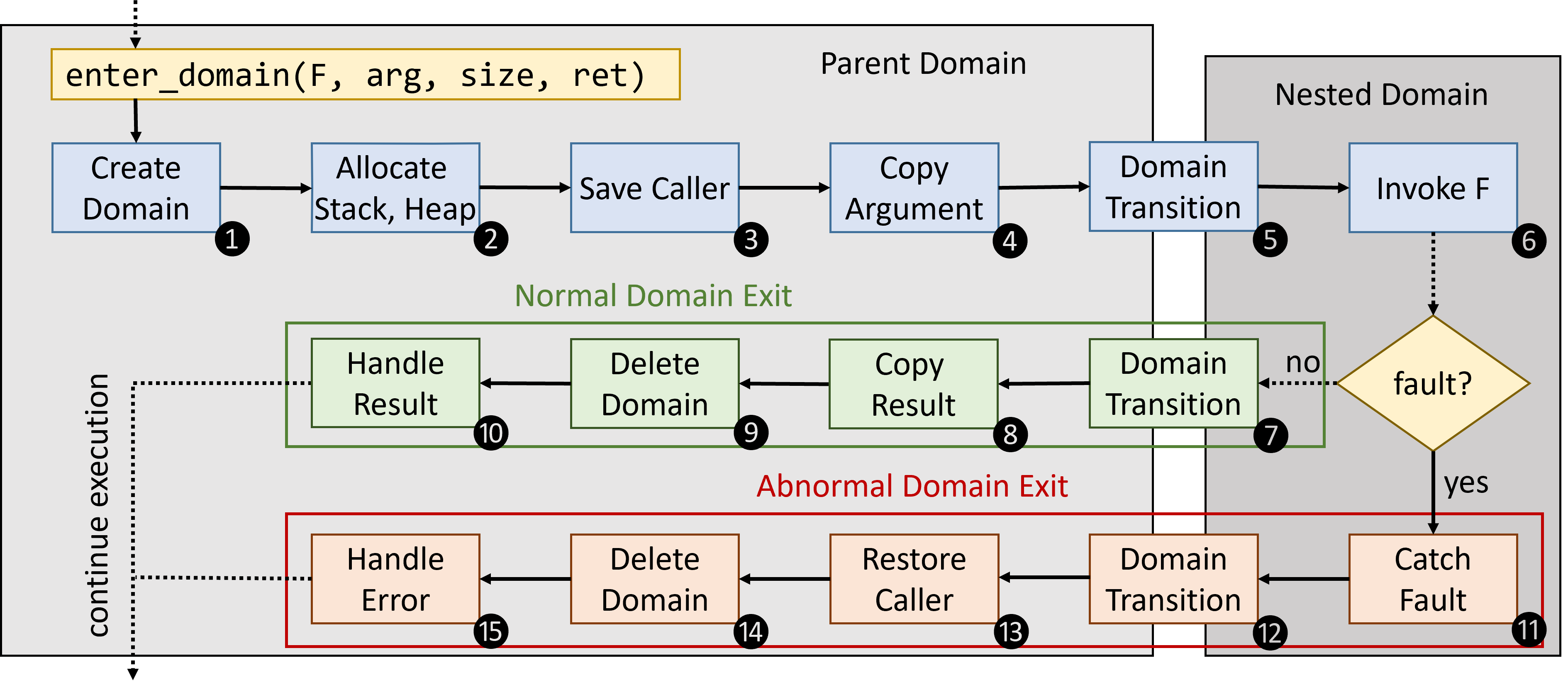} \caption{Domain
  life cycle for calling an internal or library function \emph{F} in a nested
  domain from a parent domain. Dotted arrows represent execution of user space
  instructions. Steps \protect\unicONE{} to \protect\unicTHREE{} initialize the domain. An
  argument \emph{arg} of size \protect\emph{size} is copied into the nested domain in
  step \protect\unicFOUR{} and the result is returned in variable \protect\emph{ret} in
  step \protect\unicEIGHT. Deleting the domain in steps \protect\unicNINE{} and \protect\unicFOURTEEN{}
  reverses the initialization and frees all of the nested domain's memory. Code
  for steps \protect\unicTEN{} and \protect\unicFIFTEEN{} is provided by the programmer.}\protect\label{fig:sdroblifecycle}
\end{figure*}

The overall life cycle of a nested domain in the transient style is depicted
in~\Cref{fig:sdroblifecycle}, highlighting the different steps the rollback
mechanism performs to isolate execution in a nested domain from its parent. As an example, we consider a call to a function or library \emph{F} that
takes one in-memory argument and returns a value. The call \unicSIX{} is wrapped
by \texttt{enter\_domain()}, which creates a new domain \unicONE, allocates
separate stack and heap memory \unicTWO, saves the caller context
\unicTHREE, and copies the input argument onto the new heap \unicFOUR.  The
domain transition step \unicFIVE{} performs mainly two actions:
\begin{inparaenum}[1)]
  \item reconfiguring the memory protection mechanism to restrict or grant
    memory access policies for the entered domain, \item switch between the
    stacks of the two domains.
\end{inparaenum} 
If a fault is caught in the nested domain \unicELEVEN, the abnormal domain
exit is triggered (\unicTWELVE-\unicFOURTEEN), discarding the contents of the faulty domain, and executing
the custom error handling code \unicFIFTEEN. The normal domain exit (\unicSEVEN-\unicNINE) stores the result and deletes the domain including freeing its memory.  In the ``Handle Result'' step \unicTEN, control is transferred to the developer-provided handler code for normal exits. In both exit cases, after the handler code is executed, regular program execution resumes in the parent domain. \begin{changed} For persistent domains, creation (\unicONE) is only needed for the first invocation and deletion steps (\unicNINE,\unicFOURTEEN) are omitted.\end{changed}

\subsection{Domain Nesting and Rollback} \label{sec:domainnesting}

\begin{figure}[t]
  \centering
  \includegraphics[width=0.4\textwidth]{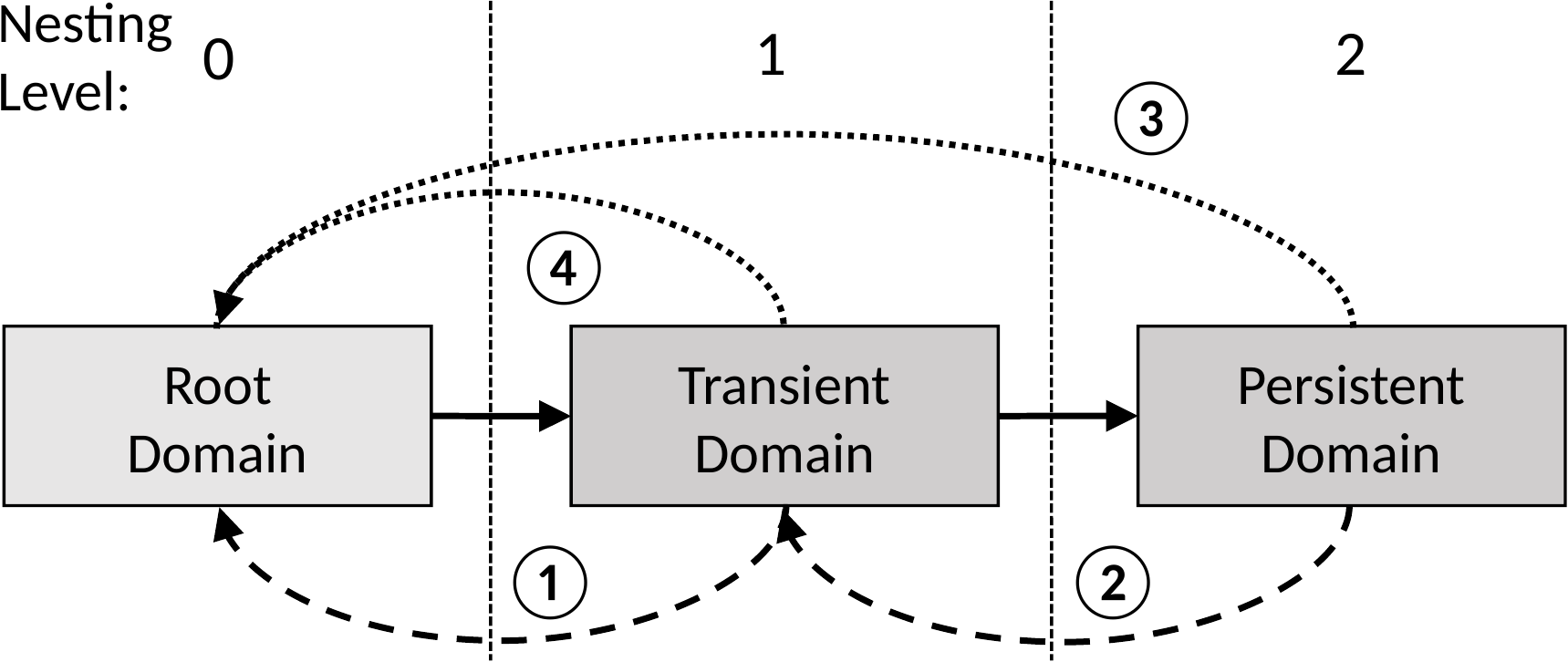}
  \caption{Deeply nested domains. Arrows indicate:
  \protect\includegraphics[scale=0.3]{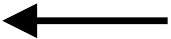} entering domain, 
  \protect\includegraphics[scale=0.3]{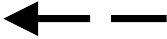} normal domain exit,
  \protect\includegraphics[scale=0.3]{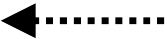} abnormal domain exit.
  Normal exits \dCTwo, \dCOne occur in reverse domain entering order. Abnormal exits \dCThree, \dCFour may deviate from that: both persistent and transient domain roll back to the \rootdomain.}
  \protect\label{fig:sdrobrollback}
\end{figure}

Domain nesting enables creating new isolated domains within others. Each nested
domain has exactly one parent domain, which is responsible for creating the
nested domain.
All domains may have zero or more nested child domains, i.e., nested
domains may be created by already nested domains.

Transient and persistent-style domains can be nested with each other. One
example of a domain nesting configuration is illustrated in
\Cref{fig:sdrobrollback}. Here, the first level of domains is
transient and its subsequent nested domain can be persistent. \begin{changed}
This setup allows developers to simplify error handling by directing rollbacks
from the more deeply nested persistent domain to also return to the recovery
point established for the transient domain. At the event of an abnormal domain
exit, the rollback can occur from any nesting level to a lower nesting level
according to application requirement%
\ifabridged, but an abnormal \rootdomain exit terminates the program.\fi
\ifnotabridged, i.e., the
rollback can be configured to occur from nesting level 2 to 1, or from 2 to 0 as
seen in \Cref{fig:sdrobrollback}. Rollbacks may occur from any nested domain,
but not from the \rootdomain. \fi
\end{changed}

\subsection{Multithreading}\label{sec:multithreading}
The secure rollback mechanism supports POSIX threads. In most cases, threads
need to communicate with each other using shared memory, hence they need to
access \rootdomain memory. Consequently, it would not be possible to
isolate two threads completely from each other or from the main process.

Nevertheless, it is still possible to isolate partial code paths within the
thread to separate domains, i.e., each thread can still create nested domains
with stacks and heaps that are isolated from the \rootdomain and other nested
domains. Hence, each thread may recover via rollback from errors in such nested
domains. If one of the threads suffers an abnormal exit from the \rootdomain, 
the rollback mechanism cannot recover other threads
and the application must be terminated.

Threads have shared access to global and root domain heap memory, to per-thread
stack areas and thread-local storage. It is possible to strengthen the isolation
by configuring exclusive memory access to each thread's stack.  However, the
security benefits are arguably marginal when heap access is still shared. A
shared \rootdomain allows a higher number of parallel threads to be supported,
if the available domains provided by the underlying memory protection mechanism
is limited, as only threads that instantiate nested domains consume domain
slots. However, one could allow the developer to configure stricter, non-uniform
access privileges to other threads.

\section{Prototype Implementation}\label{sec:implementation}
%
We implement the concept of secure domain rollback as \emph{\SDRoB} -- a 
C-language Linux library for the 64-bit x86 architecture using PKU as the
underlying isolation primitive. The library provides an API to control the life
cycle of domains.

\begin{table}
    \footnotesize
    \caption{\label{Tab:sdrobapi} \SDRoB API.  udi: user domain index.}
    \resizebox{\columnwidth}{!}{
    \begin{tabular}{|l | l|  l|}
    \hline 
     API Name  &  Arguments & Description\\   
     \hline\hline
     \hspace{-.5em}\dCOne \texttt{sdrob\_init()}   & udi, options      & Initialize Domain \emph{udi}  \\
     \hline 
     \hspace{-.5em}\dCTwo \texttt{sdrob\_malloc()}  & udi, size & Allocate \emph{size} memory in domain \emph{udi}  \\  
     \hline 
     \hspace{-.5em}\dCThree \texttt{sdrob\_free()}  & udi, adr & Free memory at \emph{adr} in domain \emph{udi}  \\  
     \hline 
     \hspace{-.5em}\dCFour \texttt{sdrob\_dprotect()}&  udi, tddi,  & Set domain \emph{udi}'s access permissions \\
                                        &  PROT                & to \emph{PROT} on target data domain \emph{tddi} \\
     \hline 
     \hspace{-.5em}\dCFive \texttt{sdrob\_enter()}   & udi      & Enter Domain \emph{udi} \\
     \hline 
     \hspace{-.5em}\dCSix \texttt{sdrob\_exit()}  &  ---          & Exit Domain \emph{udi} \\
     \hline 
     \hspace{-.5em}\dCSeven \texttt{sdrob\_destroy()}  & udi, options & Destroy Domain \emph{udi}\\
     \hline     
     \hspace{-.5em}\dCEight \texttt{sdrob\_deinit()}  & udi & Delete return context of Domain \emph{udi}\\
     \hline     
     \hspace{-.5em}\dCNine \texttt{sdrob\_call()}    & udi, fun, arg,& Convenience wrapper for single \\ 
                                        & size, ret  & function calls with one argument\\
     \hline
    \end{tabular}}
\end{table}

\subsection{\SDRoB API} 
\label{sec:api}
Developers use the \SDRoB API calls shown in \Cref{Tab:sdrobapi} to
flexibly enhance
their application with a secure rollback mechanism, accounting
for the design patterns described above.

Domains are initialized by \texttt{sdrob\_init()} \dCOne where the developer
chooses a unique index to reference the domain in future API
calls. \emph{Execution} and \emph{data} domains may be created, where the latter
may hold shareable data pages but cannot execute code. For execution domains we
further distinguish domains that are \emph{\nonisolated} or \emph{\isolated}
to their parent, and whether an abnormal domain exit should be handled in the
parent or grandparent domain. A domain can only be initialized once per thread
(unless it is deinitialized or destroyed before by the programmer) and the point
of initialization for execution domains marks the execution context to which
control flow returns in case of an abnormal domain exit.

The API call's return value fulfills two roles. When the domain is first
initialized, it returns \emph{OK} on success or an error message, e.g.,
if the domain was already initialized in the current
thread. On abnormal domain exit, control flow returns another time from the
init function and the return value signifies the index of the nested domain that
failed and was configured to return to this point. This means that error
handling for abnormal domain exits needs to be defined in a case split on the
return value of the \texttt{sdrob\_init()} function.

After initialization, memory in an execution or data domain can be managed
using \texttt{sdrob\_malloc()} \dCTwo and \texttt{sdrob\_free()} \dCThree, e.g.,
to be able to pass arguments into the domain. Note that this is only allowed for
child domains of the current domain that are \nonisolated. For \isolated
domains, a shared data domain needs to be used to exchange
data. Using \texttt{sdrob\_dprotect()} \dCFour access permissions to a data
domain can be configured for child domains.

An execution domain initialized in the current domain can be entered
and exited using \texttt{sdrob\_enter()} \dCFive and \texttt{sdrob\_exit()}
\dCSix. This switches the stack and heap to the selected domain and back, and
changes the memory access permissions accordingly. Currently, the \SDRoB
prototype does not copy local variables on such domain transitions. Such
variables need to be passed via registers or heap memory.

Supporting the transient domain design pattern, child domains can be deleted
using \texttt{sdrob\_destroy()} \dCSeven, with the option to either discard the
domain's heap memory or, if \nonisolated, merge it to the current domain. The
persistent domain pattern can then be implemented by simply not destroying the
domain after exiting it, so that it can be entered again.

An important requirement is that a nested execution domain needs
to be destroyed before the function that initialized that domain
returns. Otherwise, the stored execution context to which to return to would
become invalid as it would point to a stack frame that no longer exists. To
provide more flexibility, \texttt{sdrob\_deinit()} \dCEight
allows to just discard a child domain's execution context but leave its memory
intact. Before entering the domain again, it needs to be re-initialized, setting
a new return context for abnormal exits.

Finally, we provide \texttt{sdrob\_call()} \dCNine which
implements the life cycle of \texttt{enter\_domain()} shown
in \Cref{fig:sdroblifecycle} for a function $F$ that receives a pointer to an
object in memory as input and returns an integer-sized value. To illustrate the
API, \Cref{lst:sdrob_call} shows how \texttt{sdrob\_call()} can be implemented.

\begin{lstlisting}[style=CStyle, float, belowskip=-0.8 \baselineskip, escapechar=@@, label={lst:sdrob_call}, caption={Using other API calls (orange) to implement \texttt{sdrob\_call(udi\_F,F,arg,size,ret)} (pseudocode).}]
int err = sdrob_init(udi_F, EXECUTION_DOMAIN | 
                     ACCESSIBLE | RETURN_HERE);
if (err == OK ) {
   // prepare passing return value and argument
   register int r asm ("r12");
   register void *adr asm ("r13");
   adr = sdrob_malloc(udi_F, size); 
   if (!adr && size>0) { return MALLOC_FAILED; }   
   if (size>0) { memcpy(adr, arg, size); }
   sdrob_enter(udi_F); 
   // invoke F on copy of argument and save return value
   r = F(adr);
   sdrob_exit();
   if (ret) { *ret = r; }
   sdrob_free(udi_F, adr);
   sdrob_destroy(udi_F, NO_HEAP_MERGE);
}
return err;
\end{lstlisting}

In the example, we first initialize a new \nonisolated execution domain for
function \emph{F}. If an abnormal exit occurs in that domain,
control returns here, so we save the error code in \texttt{err}. If
initialization succeeded, we allocate a local variable \texttt{r} in a register
to retrieve the return value later. We also allocate memory for the input
argument at \texttt{adr} in the new domain. Since that domain is \nonisolated,
we can copy the argument directly from the parent domain. Afterwards, we enter
the nested domain and invoke \emph{F} on the copy of the argument, saving the
return value. After exiting, we are back in the parent domain and can copy the
return value to the desired location (which is inaccessible to the nested
domain). We free all temporary memory and destroy the nested domain, freeing its
remaining memory.\footnote{Calling \texttt{sdrob\_free()} is actually
redundant here; \texttt{sdrob\_destroy()} would free this memory as well
with the \texttt{NO\_HEAP\_MERGE} option.}  Finally, we
return \texttt{OK} in case of normal domain exit, or the error code
otherwise. Users of \texttt{sdrob\_call()} can then define their own error
handling depending on this return value as shown earlier.

\begin{figure*}
\noindent\begin{minipage}{\textwidth}
\begin{lstlisting}[style=CStyle, belowskip=-0.8 \baselineskip, label={lst:openssl_wrapper}, caption={Wrapper function for \texttt{EVP\_EncryptUpdate()} that executes OpenSSL in a persistent nested domain (excerpt). Data is passed between parent and nested domain via shared data domain \texttt{OPENSSL\_DATA\_UDI}. Error handling is omitted for brevity.}]
int __wrap_EVP_EncryptUpdate(EVP_CIPHER_CTX *ctx %*\hspace{-.5em}\dOne\hspace{-.5em}*), unsigned char *out, int *outl, const unsigned char *in, int inl) {
    register evp_encrypt_update_args_t *args asm ("r12");%*\smash\footnotemark*)  // holds copied function arguments and return value
    %*$\ldots$ *)
    %*\tikzmark{tbegin}*)
    args = sdrob_malloc(OPENSSL_DATA_UDI, sizeof(evp_encrypt_update_args_t)); 
    args->ctx = ctx;                          // copy ctx from current domain to shared data domain
    args->inl = inl;                          // copy inl from current domain to shared data domain  
    %*\tikzmark{tend}*)  
    %*\tikzmark{obegin}*)
    if (out != NULL && inl >= 0) {            // inl + cipher_block_size is upper bound for yet unknown output size 
        args.out = sdrob_malloc(OPENSSL_DATA_UDI, inl + cipher_block_size%*\smash\footnotemark*));
    } else { args.out = NULL; }
    %*\tikzmark{oend}*)                                             %*\tikzmark{ocomment}*)
    %*\tikzmark{ibegin}*)
    if (in != NULL && inl >= 0) {
        args->in = sdrob_malloc(OPENSSL_DATA_UDI, (size_t)inl);
        memcpy(args->in, in, inl);            // copy in from current domain to shared data domain
    } else { args.in = NULL; }
    %*\tikzmark{iend}*)                                             %*\tikzmark{icomment}*)
    sdrob_enter(OPENSSL_UDI);                 // execute real EVP_EncryptUpdate in inaccessible domain
    args->ret = __real_EVP_EncryptUpdate(args->ctx, args->out, &(args->outl), args->in, args->inl);
    sdrob_exit();
    %*\tikzmark{cbegin}*)
    *outl = args->outl;                       // copy out outl value from shared data domain
    if (out != NULL) {                        // copy out encrypted data from shared data domain
        memcpy(out, args->out, (size_t)*outl);
    }
    %*\tikzmark{cend}*)                                             %*\tikzmark{ccomment}*)
    %*$\ldots$*)
}
\end{lstlisting}
\begin{tikzpicture}[remember picture, overlay, thick]
    \draw[decorate,decoration={brace,amplitude=5pt,mirror}] ([shift={(-4pt,-2pt)}]pic cs:tbegin) 
    -- ([shift={(-4pt,5pt)}]pic cs:tend) 
    coordinate[midway,xshift=-6pt](Btip);
    \draw[rounded corners] (Btip) -- +(0,0)  node[left,none]
    {\scriptsize \dTwo};

    \draw[decorate,decoration={brace,amplitude=5pt,mirror}] ([shift={(-4pt,-2pt)}]pic cs:obegin) 
    -- ([shift={(-4pt,5pt)}]pic cs:oend) 
    coordinate[midway,xshift=-6pt](Btip);
    \draw[->, rounded corners] (Btip) -- +(-4pt,0) |- (pic cs:ocomment) node[right,comment,thin]
    {\scriptsize \dThree Set up output buffer. Replaced with \texttt{args.out = out} if \texttt{out} points to shared data domain.};

    \draw[decorate,decoration={brace,amplitude=5pt,mirror}] ([shift={(-4pt,-2pt)}]pic cs:ibegin) 
    -- ([shift={(-4pt,6pt)}]pic cs:iend)
    coordinate[midway,xshift=-6pt](Btip);
    \draw[->, rounded corners, bend left=45] (Btip) -- +(-4pt,0) |- ([yshift=4pt]pic cs:icomment) node[right,comment,thin]
    {\scriptsize \dFour Set up input buffer. Replaced with \texttt{args.in = in} if \texttt{in} points to a readable domain.};

    \draw[decorate,decoration={brace,amplitude=5pt,mirror}] ([shift={(-4pt,-2pt)}]pic cs:cbegin) 
    -- ([shift={(-4pt,5pt)}]pic cs:cend) 
    coordinate[midway,xshift=-6pt](Btip);
    \draw[->, rounded corners] (Btip) -- +(-4pt,0) |- (pic cs:ccomment) node[right,comment,thin]
    {\scriptsize \dFive Copy out. Omitted if \texttt{out} points to a shared data domain.};
\end{tikzpicture}
\end{minipage}
\end{figure*}

\begin{changed}
\paragraph{OpenSSL}
\Cref{lst:openssl_wrapper} shows an \texttt{EVP\_EncryptUpdate()} wrapper for
OpenSSL that implements the persistent domain pattern introduced in
\Cref{sec:domainlifecycle}. Here, OpenSSL allocates its data, such as the
context (\texttt{ctx}) \dOne from a domain which is \isolated from its parent.
While the caller can hold a pointer to \texttt{ctx}, the object itself is
inaccessible to the parent domain. Arguments are copied in via a data domain \dTwo.
The wrapped function must read buffered input and write its output to its parent domain.
There are three possible design choices for passing buffer data between the respective domains: 
\begin{inparaenum}[1)] 
\item the OpenSSL domain has read-only access to the parent, i.e., it's called
from the \rootdomain; input can be read directly, but output must be copied through
the data domain used for argument passing \dThree, \dFive
\item the parent domain is inaccessible to the OpenSSL and it must copy both
input and output via the data domain used for argument passing \dThree,
\dFour, \dFive \item the parent domain is responsible for setting up a shared
data domain between the respective domains and the wrapper can access the
shared area directly via the argument pointers.    
\end{inparaenum}

%
\addtocounter{footnote}{-1}
\footnotetext{Pointer \texttt{args} is kept in a callee-saved register (\texttt{r12}) to remain accessible after \texttt{sdrob\_enter(OPENSSL\_UDI)} (line 17) changes the domain stack.}

\addtocounter{footnote}{+1}
\footnotetext{Buffer \texttt{args.out} requires room to store
\texttt{inl} + \texttt{cipher\_block\_size} bytes of data. The
\texttt{cipher\_block\_size} is known from \texttt{ctx}~\cite{opensslman}.}

This persistent domain can be combined with a transient domain as shown
in \Cref{fig:sdrobrollback} to
\begin{inparaenum}[1)]
    \item encapsulate the pointer to \texttt{ctx} within an outer domain,
    \item protect the root domain from errors in the caller, e.g., an \texttt{out} buffer of insufficient size, and
    \item simplify error handling for the OpenSSL domain.
\end{inparaenum}
\Cref{sec:openssl_example} shows a concrete usage example.
\end{changed}



\subsection{Implementation Overview}
\label{sec:implementation_overview}

In a nutshell, \SDRoB is implemented using four components:
\begin{inparaenum}[1)] 
\item a hardware mechanism for enforcing in-process memory protection,
\item an isolated \emph{monitor data domain} that houses control data for managing execution and data domains,
\item initialization code that is run at the beginning of each application that is linked to the \SDRoB library, setting up the monitor domain and memory protection, as well as 
\item trusted reference monitor code that realizes the \SDRoB API calls, having exclusive access to the monitor data domain.
\end{inparaenum}
Below we provide more details about these components.

\paragraph{Memory Protection} 
\SDRoB uses PKU protection keys (cf.~\Cref{sec:mpk}) as a hardware-assisted SFI 
mechanism to create different isolated domains within an application governed by
different memory access policies.  When a domain is created, a unique protection
key is assigned for it. At each domain transition, PKRU is updated to grant
access to memory areas as permitted for the newly entered domain, and to prevent
access to other memory areas. Our evaluation platform supports Intel PKU, hence it allows us to manage up to
15 isolated domains at a time for each process. Software abstractions for
MPK, like libmpk~\cite{Park19}, increase the number of available domains.

\paragraph{\SDRoB Control Data} 
\SDRoB stores global control data in the monitor data domain for keeping track of,
e.g., the registered domain identifiers and the protection key usage. It also
stores per-thread information for domains such as stack size, heap size, parent
domain, and memory access permissions. To support abnormal domain exits, the
currently executing domain and the saved execution contexts are stored too.

\paragraph{Initialization}  An application is compiled with the \SDRoB 
library to use the rollback mechanism. The library provides a
constructor function that is executed before \texttt{main()} to assign all
application memory to the initial isolated domain as a \rootdomain
associated with one of the PKU protection keys. It then initializes \SDRoB global
control data where the default stack and heap size for domains is configurable
through environment variables. Furthermore, it sets the \rootdomain as active
domain, to be updated at domain transitions by the reference monitor, and
finally initializes a signal handler.  For the multithreading scenario, \SDRoB has a thread constructor function as
well, that is executed before the thread start routine function to assign a
thread memory area to a domain and associate it with one of the protection keys.

\paragraph{Reference Monitor} 
The reference monitor is responsible for book-keeping of domain information
in \SDRoB control data. It performs domain initialization, domain
memory management, and secure domain transitions, including updating the memory
access policy, and saving and restoring the execution state of the
calling domain. Only the reference monitor has access to the monitor data
domain by updating the PKRU register accordingly. The monitor code is executed
using the stack of the nested domain that invoked it.

\begin{changed}\paragraph{Error Detection}
Memory access violations are generally reported to userspace software either via
\begin{inparaenum}[1)]
    \item a \texttt{SEGFAULT} signal, e.g., when a domain tries to write
        past the confines of its memory area, or
    \item calls to runtime functions inserted by instrumentation, e.g., GCC's
        stack protector calls \texttt{\_\_stack\_chk\_fail()} if a stack guard
        check fails. 
\end{inparaenum} That function is typically
provided by glibc and terminates the application\ifnotabridged. D\fi{}\ifabridged, but we replaced it with our own implementation, allowing \SDRoB to respond to stack guard violations. Moreover, d\fi{}uring process
initialization, \SDRoB sets up its own signal handler for the \texttt{SEGFAULT}
signal, where the cause for a segmentation fault is given by a signal code
(\texttt{si\_code})  \ifnotabridged available in a \texttt{siginfo\_t}
structure~\cite{sigaction} \fi provided by the runtime to the signal
handler\ifabridged~\cite{sigaction}\fi. For instance, violations of PKU access
rules are reported by \texttt{SEGV\_PKUERR} signal code.  In Linux
the \texttt{SEGFAULT} signal is always delivered to the thread that generated
it.  If the \SDRoB signal handler detects a violation that occurred in a nested
domain, it triggers an abnormal domain exit. For faults occurring in the
\rootdomain or being attributed to a cause the \SDRoB signal handler
is not prepared to handle, the process is still terminated. 
\ifnotabridged \SDRoB also provides
its own implementation of \texttt{\_\_stack\_chk\_fail()} that replaces the
glibc implementation to respond to stack guard violations. 
\fi
\end{changed}

\SDRoB can be extended to incorporate other run-time error detection
mechanisms, such as Clang CFI~\cite{cficlang}
or heap-based overflow protections (e.g., heap red zones \cite{Serebryany19}),
improving the recovery capabilities. Probabilistic and passive protections such
as ASLR hinder the exploitation of memory safety violations but cannot detect
them. Yet, our rollback mechanism is compatible with ASLR, as domains
are created at run-time.

\paragraph{Rollback} 
The secure rollback from a domain is achieved by the reference monitor saving
the execution context of the parent domain into \SDRoB control data when that
domain is initialized. \SDRoB uses a \texttt{setjmp()}-like functionality to
store the stack pointer, the instruction pointer, the values of other registers,
and the signal mask for the context to which the call to \texttt{sdrob\_init()}
returns. Note that we cannot simply call \texttt{setjmp()}
within \texttt{sdrob\_init()} because that execution context would become
invalid as soon as the initialization routine returns. On an abnormal domain
exit, the saved parent execution state is used to restore the
application’s state to the initialization point prior to entering
the nested domain by using \texttt{longjmp()}. This rollback lets the
application continue from that last secure point of execution that is now
redirected to the developer-specified error handling code. To simplify
programming under these non-local goto semantics~\cite{setjmp}, we only allow to
set the return point once per domain and thread. Moreover, the convention that a
domain needs to be destroyed or deinitialized before the function that
initialized it returns, ensures that the saved execution context is always
valid.
\subsection{Memory Management and Isolation}\label{sec:sdrobisolation}
The secure rollback mechanism is achieved by creating different domains within
an application and ensuring that a memory defect within a domain only affects
that domain's memory, not the memory of others. Using the underlying SFI
mechanism based on PKU, each domain is isolated from other domains. \\
\paragraph{Global Variables}
We modify the linker script to ensure that global variables are allocated in a
page-aligned memory region that can be protected by PKU.
At application initialization, all global variables are assigned to the
\rootdomain; consequently they are not accessible to nested domains. As a pragmatic
solution to the problem, we make the \rootdomain by default read-only
for all nested domains. Write access to global data may then be achieved by
allocating it on the heap of a shared data domain, referenced by a global
pointer.  Note that this approach breaks the confidentiality of the \rootdomain
towards nested domains. As our main goal is integrity, and confidential data can
still be stored and processed in separate domains, we find it a reasonable
trade-off for our prototype. \\
\paragraph{Stack Management}
\SDRoB creates a disjoint stack for each execution domain to ensure that the code
running in a nested domain cannot affect the stacks of other domains. The stack
area is allocated when first initializing a domain and protected using the
protection key assigned to that domain. As an optimization, we never unmap the
stack area, even when the domain is destroyed, but keep it for reuse, i.e., when
a new domain is initialized. At each domain entry, we change the stack pointer
to the nested domain stack pointer and push the return address of
the \texttt{sdrob\_enter()} call, so that the API call returns to the call site
using the new stack. Then we update the PKRU register according to memory access
policy for that domain. A similar maneuver is performed when switching back to
the parent domain's stack via \texttt{sdrob\_exit()}. \\
\paragraph{Heap Management}
\ifnotabridged
In order to manage a domain's heap allocations as described
in \Cref{sec:domainlifecycle} the underlying allocator must have the
ability to differentiate between allocations that occur in different domains and
ensure that the underlying memory is chosen from an address range in a
particular domain's reserved heap area.  Traditionally the heap is set up to be
one large continuous memory area. Modern \texttt{malloc()}
implementations, including the GNU Allocator in
glibc~\cite{glibc} have the ability to maintain multiple disjoint heap areas, typically
for the purpose of optimizing memory access patterns in multi-threaded
applications. For example, the GNU Allocator internally maintains one or more
memory areas, referred to as \emph{arenas}, that are reserved
via \texttt{mmap()} to provide the backing memory for the application's initial
heap and subsequently allocated thread heaps.  Concurrent allocations require
threads to obtain a lock on the arena structure that \texttt{malloc()} operates
on.  Consequently, by assigning different arenas to different threads the GNU
Allocator enables memory allocations in different threads to occur concurrently
without interfering with each other.  Unfortunately the GNU Allocator's design
does not guarantee thread-isolation between arenas; if a thread fails to
allocate memory from the arena attached to it, the \texttt{malloc()}
implementation continues the search for a suitable large block of memory to
satisfy the allocation from the application's other
arenas~\cite{glibcmalloc}.
\fi
%
%
Because heap isolation in \SDRoB requires memory management with strict
guarantees that allocations within a domain are satisfied only from memory
reserved for that domain, we opted to use an allocator that natively supports
fully disjoint heap areas instead of the default glibc GNU
Allocator\ifabridged~\cite{glibcmalloc}\fi.
For our 
implementation, we chose the Two-Level Segregated Fit
(TLSF)~\cite{tlsf} allocator~\cite{mattcontetlsf}.
TLSF is a "good-fit", constant-time allocator that allocates memory blocks from
one or more pools of memory. \ifnotabridged Each free block in a pool is linked in two
different doubly linked lists:
\begin{inparaenum}[1)]
\item a free list of blocks belonging to the same size class, and
\item a list ordered by physical address.
\end{inparaenum}
The TLSF control structure contains a \emph{free list bitmap} that describes the
availability of free memory blocks in different size ranges.  TLSF uses
processor bit instructions and the bitmap to locate a corresponding linked list
of suitable-sized free blocks. \fi \\
Each \SDRoB domain is assigned its own TLSF control structure and memory pool that correspond to the domain's subheap.
The size of the initial pool assigned to domains is configurable via an environmental variable.
\ifnotabridged
Each individual TLSF pool is limited to 4GB in size~\cite{tlsf}.
Beyond that, a domain's memory is increased by reserving additional pools for the domain's TLSF allocator.
\fi 
\ifabridged
\begin{changed}
We interpose functions from the \texttt{malloc()} family with wrappers and place
the SDRoB library before libc in library load order. Upon first call to memory
management within a domain, its heap is initialized and the memory pool is
associated with the domain's protection key.  Having independent subheaps allows
the developer to either discard or merge a domain's subheap with the parent's
when the former domain is destroyed. We extended TLSF with a straight-forward
implementation of subheap merging but omit a detailed description for brevity.
\end{changed}
\fi
\ifnotabridged
\paragraph{Lazy Heap Initialization}
In order to reduce the memory footprint when domains don't make heap allocations,
the control structure and initial memory pool are not initialized when the
domain is created.  Instead, the reference monitor initializes the domain's TLSF
instance the first time the domain allocates heap memory with
the \texttt{malloc()} family of functions.
We interpose such functions with wrappers by placing the SDRoB library before
libc in the library load order.  Upon initialization the associated memory pool
is protected by associating it with the domain's protection key.
\paragraph{Subheap Merging} 
Recall from \Cref{sec:api} that a domain's subheap is either discarded or merged with
the parent domain's subheap when \texttt{sdrob\_destroy()} is called, depending on
the option parameter provided.  To enable subheap merging we extended the TLSF
implementation to associate a partially consumed pool to
a pre-existing TLSF control structure.  When a subheap is merged, its
associated protection key is updated to match the parent domain's protection
key. The parent's TLSF control structure is then updated as follows:
\begin{inparaenum}[1)]
  \item detect used and unused blocks in the memory pool being merged, 
  \item link all these blocks to the pre-existing block lists in the parent domain,
  \item update the parent domain's TLSF free list bitmap for unused blocks, and
  \item delete the child domain's TLSF control structure.
\end{inparaenum}
If the heap is not be merged when the domain is destroyed any allocations are
freed and the underlying memory pools can be reused by new domains.
\fi

\section{Case Studies}\label{sec:case_study}
\label{sec:performance}
We evaluate the performance of \SDRoB with three different
real-world case studies: Memcached, NGINX, and OpenSSL. Two aspects are
evaluated:
\begin{inparaenum}[1)]
  \item rollback latency on an abnormal exit, 
  \item performance impact of the isolation mechanism.
\end{inparaenum}
We run our experiments on Dell PowerEdge R540 machines with 24-core MPK-enabled
Intel(R) Xeon(R) Silver 4116 CPU (2.10GHz) having 128 GB RAM and using Ubuntu
18.04, Linux Kernel 4.15.0. We compiled Memcached and NGINX with \texttt{-O2}
optimizations, \texttt{-pie} (for ASLR), \texttt{-fstack-protector-strong},
and \texttt{-fcf-protection}.

\subsection{Memcached}\label{sec:memcached}
Memcached~\cite{memcached} is a general-purpose
distributed memory caching system, which is used to speed up
database-driven applications by caching database content. To do so efficiently,
Memcached stores its state in non-persistent memory; after termination and
restart, clients must start over and resend a large amount of requests to return
to the situation prior to the restart. Even in real-world deployments with
built-in redundancy and automatic remediation, small outages can take up to
a few minutes to re-route requests to an unaffected cluster~\cite{Nishtala13}.
Several studies propose to use low latency persistent
storage for Memcached~\cite{persistentmemcached,NVM_memcached} but these
solutions come with a non-negligible performance overhead. As availability and
resilience of Memcached to unforeseen failures is of high importance, it is a
worthwhile target for hardening with secure domain rollback.
%
The main thread in Memcached accepts connections and
dispatches them among worker threads to handle related requests. Memcached uses
a hash table to map keys to an index and slab allocation to manage the
in-memory database. Memcached has an event-driven architecture, handling
each client request as an event. The clients can send get, set, and update
commands with key and value arguments. To handle a request, command
parser subroutines in Memcached classify the client request, then the
key-value pairs are fetched from, inserted in, or updated in the database,
according to the client's command.  
If a client event contains a malicious request leading to memory corruption, the
database and hash table, as well as the complete application memory
area, are corrupted and Memcached must be restarted. As a result, one
malicious request affects the availability of the caching service to all clients.

\begin{figure}[t]
  \centering
  \hspace*{-0pt}\includegraphics[width=\columnwidth]{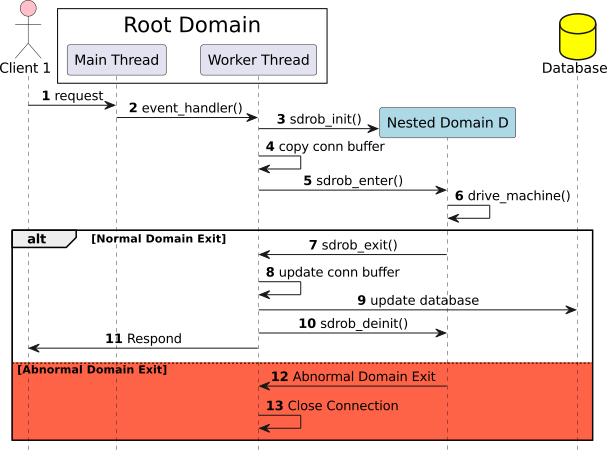}
  \caption{Sequence diagram of Memcached with \SDRoB}
  \label{fig:seq_diagram}
\end{figure}

\paragraph{Memcached with \SDRoB}\label{sec:memcached_sdrob}
We propose that each client event should be handled in a nested domain. In case
of memory corruption, the abnormal domain exit occurs in the nested domain, we
discard the related nested domain contents and come back to the \rootdomain
securely. Memcached closes the related connection, and it can continue its
execution, handling another client request without restarting. \Cref{fig:seq_diagram} shows a sequence diagram of Memcached with \SDRoB. 
We configure \SDRoB with a partially isolated multithreading configuration (see
in \Cref{sec:multithreading}), because the main thread needs to communicate with
the worker threads.  Each event is handled using the \texttt{drive\_machine()} (\textbf{6})
function with a corresponding connection buffer. We isolate this function using
the \SDRoB API, along the lines of \Cref{lst:sdrob_call}.  Recall
from \Cref{sec:domainlifecycle} that nested domains only have read
access to data that belongs to the parent domain.  Nevertheless, certain
subroutines, such as \texttt{drive\_machine()}, need to update shared state
residing in a parent domain, e.g., the connection buffer. As a solution, the
event handler that calls
\texttt{drive\_machine()} initializes an \nonisolated, nested domain $D$
(\textbf{3}) and makes a \emph{deep copy of the connection buffer} that is made
available to \texttt{drive\_machine()} (\textbf{4}). It then enters $D$
(\textbf{5}) and calls \texttt{drive\_machine()} (\textbf{6}) to handle the
client request, working on a copy of the connection buffer. After successfully
handling the request, it exits from $D$ (\textbf{7}), the original connection
buffer in the parent domain is updated with any changes present in the shared
copy (\textbf{8}). On abnormal domain exit, the copied connection buffer is
discarded.  Since the event handler returns after handling the request, we need
to invalidate the saved execution context of $D$. As \texttt{drive\_machine()}
does not allocate any persistent state in $D$, we could use the transient domain
pattern and destroy $D$.  However, as an optimization, we retain the copy of the
connection buffer used by the domain, hence \texttt{sdrob\_deinit()}
(\textbf{10}) is used.

\begin{changed}
The \texttt{drive\_machine()} function also needs to read and write the hash
table and database to perform look-ups, insertions, and updates. We allocate it
in a dedicated data domain, accessible by the nested
domain of each thread.%
\end{changed}
To allow inserts and updates, we wrap the \texttt{slabs\_alloc()}
function, that normally returns a pointer to a memory area in the database, to
return a copy of that area to insert the key-value pair. Similarly, we
wrap \texttt{store\_item()} which stores new data and updates the
hash table. Each event handler first performs its operation on a copy of the
corresponding item. On normal domain exits from $D$, we insert the
key-value pair to the database, and update the hash table (\textbf{9}). On an
abnormal domain exit (\textbf{11-12}), the corrupt key-value pair is discarded
along with all other memory of the domain. Note that this solution delays
updates to the database. However, due to the atomic nature of the Memcached
requests, consistency is not affected.
\begin{changed}
Our changes were limited to two source files in Memcached and 484 new lines of wrapper code.
In total, the changes amounted to \textasciitilde550 LoC of the 29K SLoC code base (\textasciitilde$2\%$).
\end{changed}

Memcached uses a shared mutex to synchronize worker threads. Here, our copying
mechanism for shared data does not work, because it would hide concurrent
accesses to the mutex and break the synchronization. We opted to create a
separate data domain for the mutex that every worker can
access. See \Cref{sec:applicability} for a security discussion of this scheme.

\paragraph{Rollback Latency}
We reproduced CVE-2011-4971~\cite{CVE} to verify the \SDRoB rollback mechanism
and compiled Memcached v1.4.5 with \SDRoB.  This CVE causes denial of service by
crashing Memcached via a large body length value in a packet, creating a heap
overflow but \SDRoB ensures that this overflow is limited to current execution
domain, hence it triggers the domain violation and an abnormal domain exit
occurs.  We measured the latency of abnormal domain exit starting with
catching \texttt{SEGFAULT} until after we close the corresponding
connection. The mean latency is $3.5\mathit{\mu s}$
($\sigma$=$0.9\mathit{\mu{}s}$). For comparison, in our experiments the restart and loading time for 10GiB of
data into Memcached was about 2 minutes. Thus, an attacker who successfully
launches repeated attacks could knock out the Memcached service without
rollback. While this is clearly dominated by the loading time, even
applications without such volatile state, but ultra-reliable low-latency
requirements, can benefit from rollback. For reference, we measured the mean
latency to restart the Memcached container automatically at about
$0.4s$ ($400000\mathit{\mu s}$, $\sigma$={$19000\mathit{\mu{}s}$).

\paragraph{Performance Impact}
We used the Yahoo! Cloud Service Benchmark (YCSB)~\cite{YCSB} to test the impact
of \SDRoB on Memcached performance.  YCSB has two phases: a loading phase that
populates the database with key-value pairs, and a running phase which perform
read and update operations on this data. We used workloads with sizes of 1KiB,
with a read/write distributions of 95/5. For our measurements, we stored $1
\times 10^7$ key-value pairs (1KiB each) and performed $1 \times 10^8$
operations on those pairs. Operations were performed with a Zipfian
distribution over the keys. We compiled Memcached v1.6.13 and evaluated the performance with the TLSF allocator and
with \SDRoB as described in \Cref{sec:memcached_sdrob}. We compare
the results against YCSB on unmodified Memcached.  \Cref{fig:performance} shows
the load and running phase throughput (operations/second) of the three versions
for 1, 2, 4, and 8 workers over 5 benchmark runs.  Each thread was
pinned to separate CPU cores. We used 32 YCSB clients with 16 threads pinned to
separate cores for each test.  We fully saturated Memcached cores
for 1, 2, and 4 threads but were unable to reach saturation
for 8 threads. We concluded that TLSF has negligible impact on throughput in all
our tests (<1\%).  For Memcached augmented with \SDRoB the load and running
phase overhead is $2.9\%$ / $4.1\%$, respectively, for 4 threads, and $4.5\%$ /
$5.5\%$ for 2 threads. \SDRoB introduced a worst-case overhead of $7.0\%$ /
$7.1\%$ for a single thread. We measured a performance degradation of $<4.1\%$ for
8 threads but lack confidence in the soundness of that result as the CPU was not
saturated (cf.~\Cref{tbl:throughput} in \Cref{app:throughput} for all
figures).
We measured the memory overhead of \SDRoB from the maximum resident set size
(RSS) after the YCSB load phase and comparing the RSS of Memcached with \SDRoB to
the baseline. The mean RSS increase is $0.4\%$ ($\sigma$=171KiB).

\begin{figure}[t]
    \centering
    \includegraphics[scale=0.300]{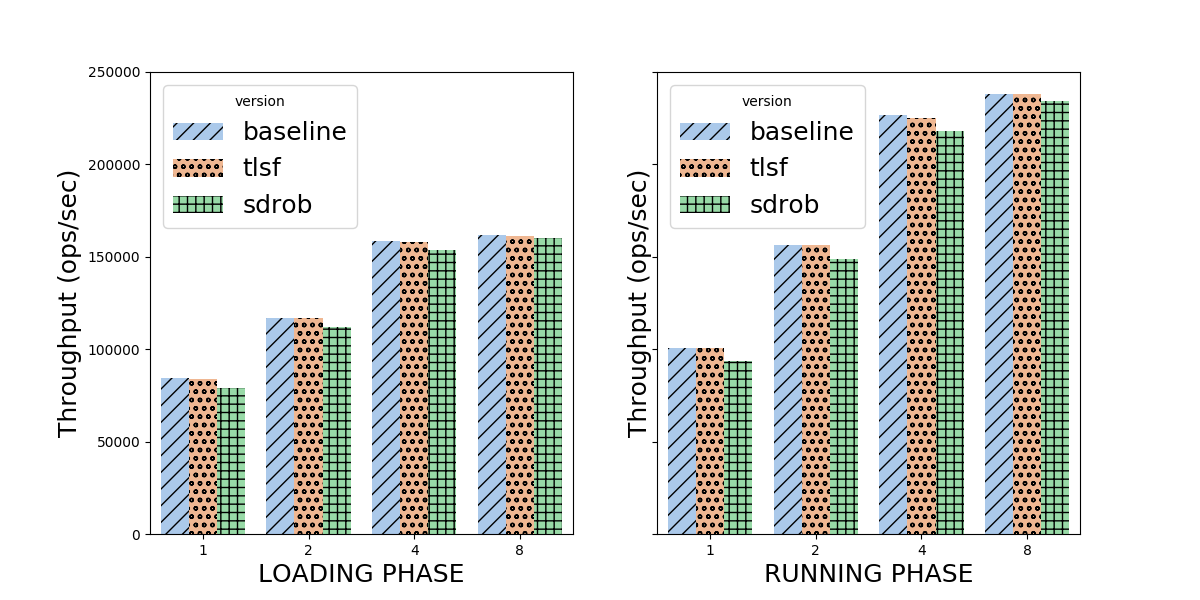}
    \caption{Throughput of different Memcached instrumentations for different numbers of threads.}
    \label{fig:performance}
\end{figure}

\begin{changed} 
\subsection{NGINX}\label{sec:nginx}
NGINX~\cite{nginx} is an open-source web server implemented as a multiprocessing
application with a master process and one or more worker processes. The master
process is responsible for maintaining the worker processes that handle client
HTTP requests for several connections at a time. If a malicious client request
leads to memory corruption, the worker process may crash and the master process
restarts it, however all active connections of that worker are lost.  Due to its
complexity and exposure to untrusted inputs, the HTTP parser is a vulnerable
component of NGINX. Similar to~\cite{endokernel}, we propose that each client
HTTP request is parsed in a nested domain. Thus, if a memory corruption is
detected in the parser, an abnormal domain exits occurs and we discard
the related nested domain content to come back to the root domain securely
without restarting the worker process. The related connection is closed, but
all other connections are unaffected.

We sandboxed the HTTP parser (ngx\_http\_parser.c) to execute in an
\nonisolated permanent nested domain, by instrumenting all NGINX parser
functions using the SDRoB API. NGINX creates a temporary memory pool for each
client request to hold a \emph{request buffer}.
We direct the allocation of these pools to a separate data domain that is
accessible by the nested domain. The request buffer data structure links back to
header data and URI data in the connection buffers. To protect this root domain
data and make it accessible to the parser, it is copied into the nested domain
and the results are copied back on domain exit. The NGINX Parser executes
multiple phases, e.g., parsing request lines or headers. Thus, domain
transitions occur repeatedly in one request. On an abnormal domain
exit, it is not important which parser phase has corrupted the memory: we always
close the corresponding connection. Hence we save as execution context first
entry point of the NGINX parser to come back to at an abnormal domain exit.
Our changes were limited to one file in NGINX and 195 new lines of wrapper code.
In total, the changes amount to \textasciitilde220 LoC of the 150K SLoC code base ($0.15\%$).

\paragraph{Rollback Latency} We reproduced CVE-2009-2629~\cite{CVE-2009-2629} to verify
the rollback mechanism and compiled NGINX v.0.6.39 with SDROB. The CVE causes a
buffer underflow in the linked connection buffer data. By having the parser
operate on copies of that data in the nested domain, the underflow triggers a
domain violation and thus an abnormal domain exit. We measured the latency of
the abnormal domain exit starting from catching SEGFAULT to accepting a new
connection. The mean latency is $3.4\mathit{\mu s}$
($\sigma$=$0.67\mathit{\mu{}s}$). We compared it with restarting the worker
process by the master process for reference.  The mean latency is
$996\mathit{\mu s}$ ($\sigma$=$44\mathit{\mu{}s}$).

\paragraph{Performance Impact} We measured the \SDRoB{} overhead to connect to NGINX remotely 
over keep-alive HTTP connections using ApacheBench (ab) tool. Each test
has 75 concurrent connections and all clients request the same file size ranging
from 0KiB to 128KiB. \Cref{fig:performance} shows mean throughput
(requests/second) of the three versions of NGINX with one worker process for
different file sizes over 5 benchmark runs.  We compiled NGINX v.1.23.1 and
compared it to NGINX with TLSF allocator and \SDRoB. The latter introduced overheads between $1.6\%$ (128KiB) and $6.5\%$ (1KiB).
We scaled up the number of workers for NGINX with \SDRoB and observed that the
overhead is independent of that number as
expected. (cf.~\Cref{tbl:nginx_throughput} in \Cref{app:throughput} for all
figures). We measured the memory overhead of \SDRoB from the maximum resident
set size (RSS) after benchmarking the 128-KiB file size with four worker
processes, and comparing the RSS of NGINX with \SDRoB to the baseline. The
mean RSS increase is $3.06\%$ ($\sigma$=50KiB).  We profiled the cost of domain
switching for NGINX and observed that $30\%-50\%$ of that cost comes from
writing to the PKRU register, which flushes the processor
pipeline~\cite{Vahldiek-Oberwagner19, Park19}.
\end{changed}

\begin{figure}[t]
  \centering
  \includegraphics[scale=0.350]{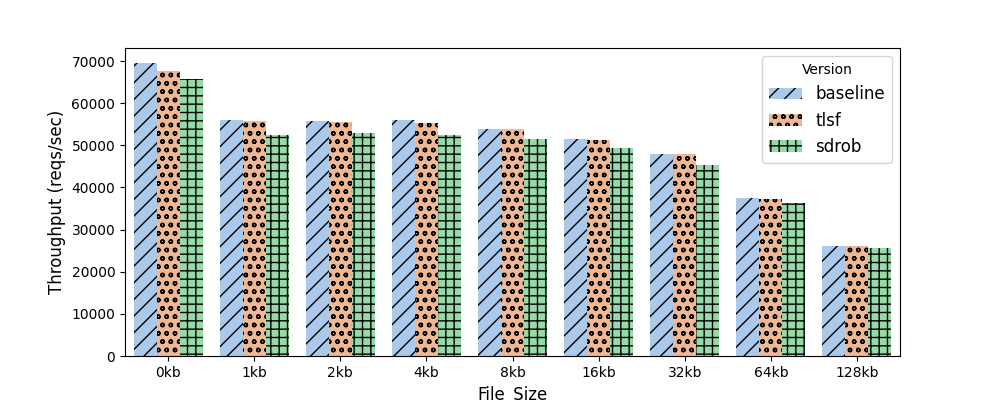}
  \caption{Throughput of different NGINX instrumentations with one worker for different file sizes.} 
  \label{fig:performance_nginx}
\end{figure}



\begin{changed} 
\subsection{OpenSSL}\label{sec:ssl}

\SDRoB allows isolating a library without changing it, enabling later
integration into applications. We evaluated the performance impact on OpenSSL~1.1.0 for all three design choices explained in \Cref{sec:api} by adapting
the built-in OpenSSL speed benchmark and running the \emph{aes-256-gcm} cipher
via the
\texttt{EVP\_EncryptUpdate} function (cf.~\Cref{lst:openssl_wrapper}) for 3s,
measuring the number of encryptions. As expected, \texttt{memcpy} operations
cause notable performance overhead and the third option, a parent-managed shared
domain, performed best. Even without copy operations, SDRoB substantially
degraded the performance of cryptographic operations for small input sizes (4\%
to 80\%).  For more realistic input sizes $\geq$32KiB we did not measure any
statistically significant overhead (< 2\%, cf.~\Cref{tbl:openssl_throughput} in
\Cref{app:throughput} for all figures).

\end{changed}

\section{Security Evaluation}\label{sec:security_ev}\label{securitydiscussion}
The primary security requirement for our work is defined with
\begin{enumerate}[label=\textbf{R{\arabic*}}, leftmargin=*]
  \setcounter{enumi}{0} \item \reqRecovery{}
\end{enumerate}
Our proposal satisfies this requirements by compartmentalizing applications
into isolated domains where an attack against a child domain can be
detected, which leads to the termination of that domain, while the parent
domain is informed and can continue operation. With respect to attack
detection, we assume that an attack or fault will exhibit an illegal memory
access that triggers a \texttt{SEGFAULT} signal. We then use signal
handlers to detect and handle the failure, leading to a termination of the
crashed child domain and a rollback of the parent domain to a well-defined
state. Compartmentalization is achieved by implementing the following two
requirements:
\begin{enumerate}[label=\textbf{R{\arabic*}}, leftmargin=*]
  \setcounter{enumi}{1} \item \reqMemoryIntegrityA{}
  \item \reqMemoryIntegrityC{}.
\end{enumerate}
In our implementation of \SDRoB, we use PKU as a mechanism to enforce
in-process isolation while facilitating efficient domain switches. The
security of this mechanism critically relies on protecting potential
gadgets in the \SDRoB implementation that allow an attacker to manipulate
the \texttt{PKRU} register.

To guarantee the security of \SDRoB, the following orthogonal defenses need
to be in place:
\begin{inparaenum}[1)]
  \item PKU crucially relies on untrusted domains to not contain unsafe
\texttt{WRPKRU} or \texttt{XRSTOR} instructions that manipulate the
\texttt{PKRU} register~\cite{Connor20}. This can be guaranteed through
W$\oplus$X and binary inspection~\cite{Vahldiek-Oberwagner19}. For programs
that do not rely on dynamic code generation, this policy can be implemented
with very low run-time overhead. Alternative proposals of hardware designs
for PKU-like security features restrict access to userspace configuration
registers~\cite{Schrammel20}.
  \item Since the \SDRoB library necessarily contains \texttt{WRPKRU}
instructions, we must employ a CFI mechanism to protect the API
implementation of our Reference Monitor and statically ensure that the
Reference Monitor API does not contain abusable \texttt{WRPKRU} or
\texttt{XRSTOR} gadgets.
  \item An alternative way to modify \texttt{PKRU} is by utilizing
\texttt{sigreturn} is described in~\cite{Connor20}: An untrusted domain may
exploit \texttt{sigreturn} gadgets and a fabricated \texttt{sigframe} crafted on
  the stack to make the kernel write an arbitrary value to the \texttt{PKRU}.
  This can happen without using a \texttt{WRPKRU} or \texttt{XRSTOR} gadget in
  userspace.  \texttt{sigreturn} attacks are mitigated by
  ASLR~\cite{lwn-sigreturn} but precluding them requires kernel-level
  authentication of \texttt{sigframe} data~\cite{Liljestrand21}.  
  \item As highlighted in~\cite{Connor20,Schrammel22}, existing PKU sandboxes do
    not sufficiently safeguard the syscall interface. Attackers can use a number
    of unsafe system calls that do not honor PKRU to erase or manipulate
    protected memory pages. Previous work \cite{Voulimeneas22,Schrammel22}
    proposes efficient syscall filtering mechanisms to prevent untrusted domains
    from invoking unsafe syscalls.
\end{inparaenum}
With the above security mechanisms in place, \SDRoB satisfies our fourth
requirement:
\begin{enumerate}[label=\textbf{R{\arabic*}}, leftmargin=*]
  \setcounter{enumi}{3} \item \reqIsolationMaintained{}
\end{enumerate}

It is possible to implement secure rollback of isolated domains on top of other isolation mechanisms
e.g., within Intel SGX to equip enclaves with rollback or by using
capability-based enforcement of isolation, e.g., 
CHERI or ARM Morello. Such uses will incur different low-level security requirements and exhibit
different performance characteristics. Furthermore, \SDRoB is not limited
to rely on \texttt{SEGFAULT} handling but could employ different attack
oracles that, e.g., trigger when a domain invokes an unexpected system
call.

\section{Discussion}\label{sec:discussion}\label{sec:extensions}\label{sec:applicability}
\paragraph{Applicability}
As highlighted earlier, secure rollback of isolated domains is particularly
suited for service-oriented applications that need strong availability
guarantees and may hold volatile state like client sessions, TLS connections,
or object caches. Redundancy and load balancing can minimize the impact of DoS
attacks, but loss of volatile state can still degrade service quality for
clients, which our approach mitigates.

\begin{changed}
\ifnotabridged
Of course, different applications will benefit from secure rollback in different
ways. A prime target for this mechanism are subroutines and libraries that
handle sequences of external, untrusted data, e.g., functions that perform input
validation, JavaScript engines in web browsers, database front-ends, or video,
image, and document renderers.  Such components exhibit a heightened degree of
exposure towards potential attacks since they operate directly on unsanitized
input. Thus, an application should isolate them in their own domain so that any
memory corruption is contained and execution can continue on a different path
after recovery via the rollback mechanism.

In practice, the specific setup of domains and protections will depend highly on
the architecture of individual applications. We expect retrofitting existing
applications written in unsafe programming languages in-lieu of a complete
re-write in a memory-safe language, to be a compelling use case. As such, the
design of secure rollback incorporates different options to
compartmentalization.

Furthermore, rollbacks that occur in long-running services may serve as
early warning signals of an attack campaign. The incident may be reported
to a Security Information and Event Management (SIEM) system and
appropriate action, such as blocking malicious clients in firewall rules
can be taken to shield the overall system from repeated attacks, minimizing
the impact on legitimate clients.
\fi
\ifabridged 
A prime target for this mechanism are subroutines and libraries that handle
sequences of external, untrusted, unsanitized data, e.g., functions that perform
input validation, JavaScript engines in web browsers, database front-ends, or
video, image, and document renderers, due to a heightened degree of exposure
towards potential attacks. Isolating such components in their own domain allows
to recover from potential memory corruptions. Furthermore, rollbacks in
long-running services may be reported as incidents to a Security Information and
Event Management (SIEM) system, serving as early warning signals of an attack
campaign. Blocking malicious clients via firewall rules as a response may then
shield the overall system from repeated attacks.

In practice, the specific setup of domains and protections will depend highly on
application architecture. As such, our design incorporates different options to
compartmentalization, aiding adoption of the rollback mechanism in new and
existing developments. We see retrofitting applications written in unsafe
programming languages as a compelling use case in-lieu of a complete re-write in
a memory-safe language.
\fi
\end{changed}

%

\paragraph{Limitations}
It is clear that not all applications can be easily compartmentalized and
refactored to make use of \SDRoB. For example, applications that rely on
global mutexes might suffer from availability issues when a child domain
that holds a lock crashes and the lock is not released prior to
continuation of the parent domain. Options for resolving this are, e.g., to
provide an \SDRoB-aware locking mechanism as part of our \SDRoB library, or to
\begin{changed}
prefer local locks with well-defined scope instead of global locks. This
aids serializing access, e.g., when domains operate on copies of protected
objects.
\end{changed}

Another potential issue comes with complex data structures used by
target applications. Similar to other strong isolation mechanisms such as
Intel SGX, data needs to be copied into the address space
of the protection domain~\cite{costan2016intel}, which is done by entry
wrappers. Both, manual as well as automated generation of these wrappers
can be error prone and may hamper security~\cite{van2019tale}.  Generally,
domain transitions and domain termination bear subtle risks.  Currently,
confidentiality of child domain data is not guaranteed after destroying it and we leave it to the developer to realize such requirements, e.g.,
scrub sensitive allocations from memory before leaving the domain.

The use of \SDRoB as a mechanism that increases the availability of
long-running services may open up a side-channel attack surface. As
observing errors might give an attacker insights into an
application's execution, the observable effects of a rollback (e.g.,
delayed execution) might also give such insights. Coupled with the absence
of re-randomization of the application's memory layout, an attacker could
potentially use this to break probabilistic defenses such as ASLR.
A potential protection against such attacks could be achieved by making the
rollback behavior of \SDRoB configurable and force an application restart
after a certain number of rollbacks, similarly to probabilistic defenses for
pre-forking applications~\cite{Liljestrand19a}.

Ultimately the security of \SDRoB depends on the correctness of our library
implementation and further exploration of the attack surface and
potentially formal verification of our code are envisaged to harden our
approach. 

\ifnotabridged
\begin{changed}
\ifnotabridged
Additionally, we further envision the following possible extensions to \SDRoB:

\paragraph{Global variables}
Currently, global variables are stored in the \rootdomain that is read-only by
default to allow access to all nested domains. A more comprehensive solution
would place these variables in their own dedicated data domain(s) with a
reserved identifier.

\paragraph{Accessing local variables}
As discussed in \Cref{sec:api}, having dedicated stacks per domain precludes accessing local variables across domain transitions. Beyond the use of registers for passing arguments
and heap memory in shared domains for passing data, \SDRoB could also incorporate existing compiler support for disjoint stacks present in GCC~\cite{glibcsplitstack} to facilitate accesses to objects across stack boundaries, e.g., arguments passed on the stack.

\paragraph{Access control}
At the moment, a very simple policy governs access to domains: in principle only
parent domains can perform API calls targeting a nested domain. Still, more
flexible policies may be desirable, i.e., inheriting access to data domains from
a parent. However, such features need to be designed carefully, not to
over-complicate the programming model, or to compromise security.

\paragraph{Strongly Isolated Multithreading Design}
\SDRoB currently employs the \emph{partially isolated multithreading design}
discussed in \Cref{sec:multithreading}. However, if an application employs
threads that do not need to synchronize on shared data, e.g., worker threads
processing local data, then each thread can have a separate isolated \rootdomain
containing each thread's stack and heap area. This strongly isolated setup
prevents that a compromised thread can access any data owned by other threads.
Furthermore, in case of abnormal domain exit from a thread \rootdomain, the
rollback mechanism can be configured to discard all the compromised thread's
domains and recreate the thread from scratch without aborting the
application. It would be possible to allow the setup and initialization of
thread \rootdomain{}s to the configurable by the programmer, to allow for hybrid
configurations where a subset of threads may fail without affecting the main
process. To this end we would add thread creation to the \SDRoB API to let the
programmer specify thread \rootdomain identifiers, instead of wrapping
\texttt{pthread\_create()} as is done now. This would allow allocating threads
to the same \rootdomain{}s explicitly and also to share data domains with
otherwise strongly isolated threads.

\paragraph{Thread safety}
Different threads can initialize the same domain individually. However, our
current implementation of domains is not thread-safe apart from shared
\rootdomain{}s, i.e., it is undefined what happens if two threads enter a given nested
domain at the same time. The programmer needs to ensure that entry to shared
nested domains (e.g., an isolated crypto library) is synchronized. While our
internal data structures can support concurrent domain entry, more work is
needed to manage abnormal domain exits. If such an exit is triggered by one
thread, it should be propagated automatically to all other threads that might be
running in that domain.
\fi
\ifabridged
\paragraph{Extensions}
Additionally, there are a number of currently missing features that could improve our \SDRoB prototype. For instance, global variables are currently stored in the \rootdomain with read-only access for nested domains. We could add support to place globals in a dedicated data domain. 

As discussed in \Cref{sec:api}, local variables currently do not remain valid
across domain transitions due to stack isolation. Beyond the use of registers
for passing arguments and heap memory in shared domains for passing data,
compiler support for disjoint stacks present in GCC~\cite{glibcsplitstack} could
facilitate safe accesses to local variables across domain boundaries.

\SDRoB currently employs the \emph{partially isolated multithreading design}
 with a shared \rootdomain (cf.~\Cref{sec:multithreading}). However, if application worker threads only process local data, each thread can have a separate isolated \rootdomain for its stack and heap area. This strongly isolated setup
prevents compromised threads from accessing data owned by other threads and abnormal domain exits from a thread \rootdomain do not crash the application anymore. Instead the rollback mechanism could discard the compromised thread's
domains and recreate it from scratch. A more comprehensive approach would to allow the programmer to explicitly setup and initialize thread \rootdomain{}s with chosen identifiers.

While different threads can initialize the same domain individually, domains are currently not thread-safe apart from shared \rootdomain{}s, i.e., it is undefined what happens if two threads enter a given nested domain at the same time. The programmer needs to ensure that entry to shared nested domains (e.g., an isolated crypto library) is synchronized. Our internal data structures can support concurrent domain entry, but a mechanism is needed to propagate abnormal domain exits to all other threads that might be running in that domain.

Overall, the above concerns are not fundamental limitations but mere engineering challenges in order to provide a production-ready \SDRoB solution.
\fi
\end{changed}
\fi
%
%
%
%
%

\section{Related Work}\label{sec:related_work}
\begin{changed}
\paragraph{Hardware-assisted compartmentalization}
The idea of using MPK for compartmentalizing applications is not new; PKU in
64-bit x86 is used to augment SFI approaches that generally suffer from high
enforcement overheads~\cite{Wahbe93, David_SFI}. Such work generally falls in
two broad categories:
\begin{inparaenum}[1)]
  \item in-process isolation~\cite{Rivera16,Koning17,Hedayati19,Vahldiek-Oberwagner19,Schrammel20,Wang20,Voulimeneas22,Kirth22,Jin22,Chen22}, and
  \item isolation for unikernels and library operating systems~\cite{Melara19,Sung20,Lefeuvre21}.
\end{inparaenum}
In-process isolation using PKU has been scrutinized for the lack of controls on
PKRU access which may leave schemes vulnerable to attacks that bypass
established isolation domains~\cite{Connor20,Schrammel22,Voulimeneas22}.
Countermeasures proposed against such attacks, include code
rewriting~\cite{Koning17}, binary inspection~\cite{Vahldiek-Oberwagner19},
system call filtering~\cite{Voulimeneas22,Schrammel22} and variations on the PKU
hardware design~\cite{Schrammel20,Delshadtehrani21,Frasetto18}. Secure
multi-threading has also been considered~\cite{Chen16,mutiles,Kirth22}.
However, in-process SFI, similar to the defenses discussed in
\Cref{sec:background}, does not consider how to recover from attacks
regardless of whether enforced via software or hardware. This work addresses
this gap by introducing capabilities for secure rollback.

Capability hardware, such a CHERI~\cite{Woodruff14,Watson15} also enables
compartmentalized fault isolation. CompartOS~\cite{Almatary2022} provides
recovery capabilities for CHERI faults in compartments, but unlike \SDRoB, it is
geared toward safety-critical embedded systems, not commercial off-the-shelf processors.
\end{changed}

\paragraph{Checkpoint \& restore}
Existing approaches to checkpoint \& restore, such as CRIU~\cite{Kashyap16} provide
support for process snapshots that can enable rollback-like functionality.
However, checkpoint \& restore generally suffers from high overheads due to
relying on reproducing process memory and do not consider in-memory attacks in
their threat models~\cite{optimizechekpoint, Young1974AFO, checkpointrestart,
seccheck, webster2018malware}. \SDRoB avoids these drawbacks by combining
in-process isolation to limit the scope of attacks and to ensure the integrity
of memory after rollback.  

\paragraph{N-variant Execution}
\ifabridged
\begin{changed}
N-variant Execution (NVX)~\cite{NVX} provides resilience against invasive
attacks by introducing redundancy through running multiple, artificially
diversified variants of the same application in tandem and terminating instances
that show divergent behavior. While \SDRoB shares the goal of improving software
resilience with NVX, our work targets use cases for which the high cost of
replicating computations and I/O across each instance is impractical.
\end{changed}
\fi
\ifnotabridged
N-variant Execution (NVX)~\cite{NVX} provides resilience against invasive attacks by
introducing redundancy through running multiple, artificially diversified
variants of the same application in tandem and monitor each distinct copy for
divergent behavior. If any inconsistencies between the instances are detected
NVX terminates the offending instances’ execution while unaffected instances
can continue. While \SDRoB shares the goal of improving software resilience
with NVX we consider use cases for which the high cost of replicating compute
instances and I/O across each instance is impractical. Safety-critical applications for which the high deployment cost of NVX may be
justified are outside the scope of this work.
\fi

\section{Conclusion \& Future Work}
\label{sec:conclusion}\label{sec:future_work}\label{sec:availability}
\label{futurework}

We presented the novel concept of \emph{secure rollback of isolated domains} which complements
protection mechanisms against memory safety vulnerabilities. It provides a
hardening mechanism to recover from detected violations, thus improving the availability and resilience of software applications. 
At its core, secure rollback uses hardware-assisted in-process
memory isolation to isolate exposed functionality in separate domains, so
that a compromise in one domain cannot spread to other parts of a program's
memory. When a compromise is detected by selected defense
mechanisms, a rollback to a previously defined consistent state of the
application occurs, enabling error handling and resuming the application.

We explored various design patterns for domains and presented SDRoB, our
prototype library implementation of secure rollback. We demonstrated
its applicability to real software by adding it to the multi-threaded
Memcached system, \begin{changed} multiprocessing NGINX and the OpenSSL library.\end{changed} 

Besides alleviating current limitations discussed above, we
aim to improve usability for the programmer, e.g., by providing a domain
specific language and compiler support for the definition of domains and the
exchange of data between them\ifnotabridged, similar to the edger8r tool used for Intel SGX
enclaves\fi. Providing an amenable, secure, and efficient
implementation of the secure rollback mechanism will fill an important gap in
the current software security architecture.

%


\ifnotanonymous
\section{Acknowledgments} 
We thank Ilhan G\"{u}rel, Michael Liljestam, Eddy Truyen, Sini Ruohomaa, Sava Nedeljkovic, Prajwol Kumar Nakarmi, Quentin Stievenart
Dominique Devriese, and Anjo Vahldiek-Oberwagner for their feedback, 
which helped improve the paper.
We further thank Stijn Volckaert and his team at KU Leuven -- Ghent for providing the infrastructure 
to run our experiments, and for his feedback on our work. This research is partially funded by the 
Research Fund KU Leuven, by the Flemish Research Programme Cybersecurity. 
This research has received funding under EU H2020 MSCA-ITN action 5GhOSTS, 
grant agreement no. 814035.
\fi

\ifarxiv
\bibliographystyle{ACM-Reference-Format}
\fi

\ifnotarxiv
\bibliographystyle{plainurl}
\fi
\bibliography{main}

\appendix
\clearpage
\begin{minipage}{\textwidth}
\section{Supplementary Measurements}
\label{app:throughput}

\begin{multicols*}{2}
\Cref{tbl:rollbacklatency} shows the detailed results for our rollback latency
measurements for Memcached. The \emph{Rollback latency} column gives the mean
latency in $\mathit{\mu{}s}$ over 1000 rollback iterations and the
\emph{$\sigma$} column the standard deviation. We measured both the rollback
that destroys the offending domain but leaves the contents of its memory intact
and a version of the rollback that "scrubs" (zeroes) the full contents of domain
memory before it can be re-allocated, as indicated in the \emph{Zeroing of
domain data column}. The latter estimates the upper bound for maintaining
confidentiality guarantees for nested domains using a 4GB heap pool and 4MB
domain stack thats store sensitive information as discussed in
\Cref{sec:discussion}. As rollbacks are exceptional events, we deem the $0.2s$
latency reasonable for use cases with confidentiality requirements. The latency
could be reduced by tracking and scrubbing only allocations that contain
sensitive data, or reducing the reserved memory for confidential domains.

\Cref{tbl:memoryconsumption} shows the detailed results of our memory
consumption measurements for Memcached. The \emph{Maximum resident set size}
column reports the mean maximum resident size (RSS) reported by the Bash shell
builtin \texttt{time} command over five iterations of the YCSB benchmark loading
phase for the unmodified baseline (\emph{Baseline}) and Memcached equipped with
\SDRoB (\emph{SDRoB}). The columns marked \emph{$\sigma$} give the relative
standard deviation of the RSS. We used used four worker threads for this
experiment as indicated in the \emph{\#Thr} column.

\Cref{tbl:throughput} shows the detailed result of our throughput measurements
of the YCSB benchmark for Memcached using different numbers of threads
(\emph{\#Thr}) that were summarized in \Cref{sec:performance}. The
\emph{Throughput} column gives the throughput in operations / seconds of three
versions of Memcached: {\captionenumber{ \protect\item The unmodified baseline
    (\emph{Baseline}), \protect\item Memcached using the TLSF allocator
    (\emph{TLSF}), and \protect\item Memcached equipped with \SDRoB
    (\emph{SDRoB}).}} The columns marked \emph{$\sigma$} give the relative
standard deviation of the throughput over ten benchmark runs as percentage for
the aforementioned versions. The \emph{Throughput degradation} column gives the
degradation of throughput in percentage of {\captionenumalpha{ \protect\item
    Memcached using the TLSF allocator compared to the baseline
    (\emph{TLSF/Baseline}), \protect\item Memcached equipped with \SDRoB
    compared to Memcached using the TLSF allocator (\emph{\SDRoB/TLSF}), and
    \protect\item Memcached equipped with \SDRoB compared to baseline
    (\emph{\SDRoB/Baseline}).}}

\begin{changed}
\Cref{tbl:nginx_rollbacklatency} shows the detailed results for our rollback
latency measurements for NGINX. The columns are similar to those in
\Cref{tbl:rollbacklatency}.

\Cref{tbl:nginx_memoryconsumption} shows the detailed results of our memory
consumption measurements for NGINX. We report the mean maximum RSS over 10
iterations of the NGINX benchmark for the unmodified baseline (\emph{Baseline})
and NGINX equipped with \SDRoB (\emph{SDRoB}). The columns marked
\emph{$\sigma$} give the relative standard deviation of the RSS. We used 4
workers processes for this experiment as indicated in the \emph{\#Wkr} column.

\Cref{tbl:nginx_throughput} shows the detailed result of our throughput
measurements of the ab tool for NGINX using different file sizes starting from
0KiB to 128KiB with different worker processes that were summarized in
\Cref{sec:performance}.  The \emph{Throughput} column gives the throughput in
requests / seconds of three versions of NGINX: {\captionenumber{ \protect\item
    The unmodified baseline (\emph{Baseline}), \protect\item NGINX using the
    TLSF allocator (\emph{TLSF}), and \protect\item NGINX equipped with \SDRoB
    (\emph{SDRoB}).}} The columns marked \emph{$\sigma$} give the relative
standard deviation of the throughput over five benchmark runs as percentage for
the aforementioned versions. The \emph{Throughput degradation} column gives the
degradation of throughput in percentage of {\captionenumalpha{ \protect\item
    NGINX using the TLSF allocator compared to the baseline
    (\emph{TLSF/Baseline}), \protect\item NGINX equipped with \SDRoB compared to
    NGINX using the TLSF allocator (\emph{\SDRoB/TLSF}), and \protect\item NGINX
    equipped with \SDRoB compared to baseline (\emph{\SDRoB/Baseline}).}}

\Cref{tbl:openssl_throughput} the detailed result of our throughput
measurements of the OpenSSL benchmark tool for different input sizes from 16 to 
262144 bytes that were summarized in 
\Cref{sec:performance}. The \emph{Throughput} column gives the throughput in
1000s of bytes / seconds of three versions of OpenSSL: \SDRoB 1. -- 3. correspond to the three different design choices for the \texttt{EVP\_EncryptUpdate()} wrapper described in \Cref{sec:api}: 
\SDRoB 1. -- OpenSSL domain has read-only access to parent, 
\SDRoB 2. -- the wrapper is responsible for copying input/output via an
intermediate data domain, 
\SDRoB 3. -- the parent domain sets up a shared data domain which the OpenSSL domain in the wrapper can access directly.
\end{changed}
\end{multicols*}
\end{minipage}

\begin{figure*}
\begin{minipage}[t]{\columnwidth}
\begin{table}[H]
    \begin{center}
        \caption{Memcached: Rollback latency}
        \label{tbl:rollbacklatency}
        \begin{tabular}{|l|r|r|}
            \hline 
            \multirow{2}{*}{Zeroing of domain data} & \multicolumn{2}{|c|}{Rollback latency ($\mathit{\mu{}s}$)} \\
            & \multicolumn{1}{|c|}{\SDRoB} & \multicolumn{1}{|c|}{$\sigma$} \\
            \hline\hline
            No (default) &  3.46\hspace{2.5em} & $\pm0.9\mu{}s$ \\
            \hline 
            Yes & 228376.99 ($0.2s$) & $\pm2500\mu{}s$ \\
            \hline 
        \end{tabular}
    \end{center}
\end{table}
\end{minipage}
\begin{minipage}[t]{\columnwidth}
\begin{table}[H]
    \begin{center}
        \caption{\begin{changed}NGINX: Rollback latency\end{changed}}
        \label{tbl:nginx_rollbacklatency}
        \begin{tabular}{|l|r|r|}
            \hline 
            \multirow{2}{*}{Zeroing of domain data} & \multicolumn{2}{|c|}{Rollback latency ($\mathit{\mu{}s}$)} \\
            & \multicolumn{1}{|c|}{\SDRoB} & \multicolumn{1}{|c|}{$\sigma$} \\
            \hline\hline
            No (default) &  3.41\hspace{2.5em} & $\pm0.7\mu{}s$ \\
            \hline 
            Yes & 232632,4168 ($0.2s$) & $\pm 2400\mu{}s$ \\
            \hline 
        \end{tabular}
    \end{center}
\end{table}
\end{minipage}
\end{figure*}
    
\begin{figure*}
\begin{minipage}[t]{\columnwidth}
\begin{table}[H]
    \begin{center}
        \caption{Memcached: Memory consumption}
        \label{tbl:memoryconsumption}
        \begin{tabular}{|c||c|c|c|c|c|}
            \hline 
            \multirow{2}{*}{\#Thr} & \multicolumn{4}{|c|}{Maximum resident set size (KiB)} \\
            & Baseline & $\sigma$ & \SDRoB & $\sigma$    \\
            \hline\hline
            4 & 14535444.8 & $\pm1.19\%$ &  14598972.8 & $\pm1.18\%$ \\
            \hline 
        \end{tabular}
    \end{center}
\end{table}
\end{minipage}
\begin{minipage}[t]{\columnwidth}
\begin{table}[H]
    \begin{center}
        \caption{ \begin{changed}NGINX: Memory consumption \end{changed}}
        \label{tbl:nginx_memoryconsumption}
        \begin{tabular}{|c||c|c|c|c|c|}
            \hline 
            \multirow{2}{*}{\#Wkr} & \multicolumn{4}{|c|}{Maximum resident set size (KiB)} \\
            & Baseline & $\sigma$ & \SDRoB & $\sigma$    \\
            \hline\hline
            4 & 3135.6   &  $\pm 1.59\%$ &  3234.8 & $\pm 1.71\%$ \\
            \hline 
        \end{tabular}
    \end{center}
\end{table}
\end{minipage}
\end{figure*}

\begin{figure*}
\begin{minipage}[t]{\textwidth}
\begin{table}[H]
\begin{center}
\caption{Memcached: Detailed table of throughput measurements.}
    \label{tbl:throughput}
    \begin{tabular}{|r|r|c|r|c|r|c|c|c|c|}
  \hline 
     \multirow{2}{*}{\#Thr} & \multicolumn{6}{|c|}{Throughput (ops/sec)} & \multicolumn{3}{|c|}{Throughput degradation (\%)} \\
        & Baseline & $\sigma$ & TLSF & $\sigma$ & SDRoB  & $\sigma$ & {\footnotesize TLSF/Baseline} & {\footnotesize\SDRoB/TLSF} & {\footnotesize\SDRoB/Baseline} \\
  \hline\hline
  \multicolumn{10}{|c|}{Loading Phase} \\
  \hline
  1     & 83913  & $\pm$0.69\%  & 84191  & $\pm$0.46\% & 78047  & $\pm$0.89\%  & +0,33\% & -7.30\% & -6.99 \% \\ \hline          
  2     & 117194 & $\pm$1.05\%  & 117275 & $\pm$0.18\% & 111838 & $\pm$0.40\%  & +0,07\% & -4.64\% & -4.57 \% \\ \hline          
  4     & 157920 & $\pm$1.24\%  & 157982 & $\pm$0.43\% & 153328 & $\pm$0.63\%  & +0,04\% & -2.95\% & -2.91 \% \\ \hline          
  8     & 161161 & $\pm$0.80\%  & 161448 & $\pm$0.75\% & 160910 & $\pm$1.13\%  & +0,18\% & -0.33\% & -0.16 \% \\
  \hline\hline        
  \multicolumn{10}{|c|}{Running Phase} \\\hline
  1     & 99722  & $\pm$0.46\%  & 99829  & $\pm$0,52\%  & 92645   & $\pm$0,23\%  & +0,11\% & -7,20\% & -7,10\% \\\hline
  2     & 155247 & $\pm$0.54\%  & 155015 & $\pm$0,51\%  & 146715  & $\pm$0,17\%  & -0,15\% & -5,35\% & -5,50\% \\\hline
  4     & 223958 & $\pm$0.20\%  & 223886 & $\pm$0,41\%  & 214671  & $\pm$0,54\%  & -0,03\% & -4,12\% & -4,15\% \\\hline
  8     & 234823 & $\pm$0.76\%  & 235007 & $\pm$0,55\%  & 225277  & $\pm$0,62\%  & +0,08\% & -4,14\% & -4,07\% \\ \hline
 \end{tabular}
\end{center}
\end{table}
\end{minipage}
\end{figure*}

\begin{figure*}
\begin{minipage}[t]{\textwidth}
\begin{table}[H]
    \begin{center}
        \caption{\begin{changed}OpenSSL: Detailed table of throughput measurements. \end{changed}}
            \label{tbl:openssl_throughput}
            \resizebox{\textwidth}{!}{
                \begin{tabular}{|r|r|c|r|c|r|c|c|c|c|c|c|}
                    \hline 
                    Input     & \multicolumn{8}{|c|}{Throughput (1000s of bytes/sec)} & \multicolumn{3}{|c|}{Throughput degradation (\%)} \\
                    bytes     & Baseline &  $\sigma$  &\SDRoB 1.&  $\sigma$   &\SDRoB 2.&  $\sigma$   &\SDRoB 3.&  $\sigma$   & 1./Baseline & 2./Baseline & 3./Baseline \\   
                    \hline
                    $2^4$     &  267879 & $\pm$1.04\% &   53570 & $\pm$2.08\% &   33995 & $\pm$1.41\% &  53851  & $\pm$3.02\% & -80.00\% & -87.31\% & -79.90\% \\
                    \hline                                                                                                                
                    $2^6$     &  724060 & $\pm$1.47\% &  196647 & $\pm$1.84\% &  127678 & $\pm$1.02\% &  201048 & $\pm$2.88\% & -72.84\% & -82.37\% & -72.23\% \\
                    \hline                                                                                                                
                    $2^8$     & 1508889 & $\pm$2.11\% &  616788 & $\pm$1.79\% & 1192933 & $\pm$1.32\% &  639910 & $\pm$2.07\% & -59.12\% & -71.59\% & -57.59\% \\
                    \hline                                                                                                                
                    $2^{10}$  & 2515179 & $\pm$1.85\% & 1551532 & $\pm$1.66\% & 428704  & $\pm$1.74\% & 1610508 & $\pm$1.76\% & -38.31\% & -52.57\% & -35.97\% \\   
                    \hline                                                                                                                
                    $2^{13}$  & 3073966 & $\pm$1.96\% & 2744814 & $\pm$1.17\% & 2503868 & $\pm$1.51\% & 2842073 & $\pm$1.10\% & -10.71\% & -18.55\% &  -7.54\% \\   
                    \hline                                                                                                                
                    $2^{14}$  & 3139433 & $\pm$1.78\% & 2856443 & $\pm$1.11\% & 2567001 & $\pm$1.18\% & 3025082 & $\pm$1.52\% &  -9.01\% & -18.23\% &  -3.64\% \\   
                    \hline                                                                                                                
                    $2^{15}$  & 3167069 & $\pm$1.16\% & 2855263 & $\pm$1.13\% & 2577581 & $\pm$1.48\% & 3111608 & $\pm$1.35\% &  -9.85\% & -18.61\% &  -1.75\% \\   
                    \hline                                                                                                                
                    $2^{16}$  & 3184860 & $\pm$1.18\% & 2904903 & $\pm$0.71\% & 1532381 & $\pm$1.60\% & 3169666 & $\pm$1.23\% &  -8.79\% & -51.89\% &  +0.28\% \\   
                    \hline                                                                                                                
                    $2^{18}$  & 3198742 & $\pm$1.14\% & 2941696 & $\pm$0.83\% & 2637291 & $\pm$1.54\% & 3207550 & $\pm$1.05\% &  -8.04\% & -17.55\% &  +0.38\% \\   
                    \hline
                \end{tabular}}
    \end{center}
\end{table}
\end{minipage}
\end{figure*}

\begin{figure*}
\begin{minipage}[t]{\textwidth}
\begin{table}[H]
    \begin{center}
    \caption{\begin{changed}NGINX: Detailed table of throughput measurements.\end{changed}}
        \label{tbl:nginx_throughput}
        \resizebox{\textwidth}{!}{
        \begin{tabular}{|r|r|c|r|c|r|c|c|c|c|c|}
        \hline 
            \multirow{2}{*}{\specialcell{\footnotesize File size \\ \footnotesize (KiB)}} & \multicolumn{6}{|c|}{Throughput (reqs./sec.)} & \multicolumn{3}{|c|}{Throughput degradation (\%)} \\
            & Baseline & $\sigma$ & TLSF & $\sigma$ & SDRoB  & $\sigma$ & {\footnotesize TLSF/Baseline} & {\footnotesize\SDRoB/TLSF} & {\footnotesize\SDRoB/Baseline} \\
         \hline\hline   
         \multicolumn{10}{|c|}{1 worker} \\
         \hline
      0     & 69386 & $\pm$1.40 \%  & 66953 & $\pm$2.21\% & 65025 & $\pm$1.04\%  & -2.96\% & -2.75\% & -5.63\%  \\ \hline          
      1     & 55454 & $\pm$1.01 \%  & 55484 & $\pm$0.74\% & 51839 & $\pm$1.48\%  & -0.31\% & -6.17\% & -6.46\%  \\ \hline          
      2     & 55184 & $\pm$0.81 \%  & 55273 & $\pm$0.26\% & 52482 & $\pm$0.70\%  & -0.21\% & -4.93\% & -5.13\%  \\ \hline          
      4     & 55506 & $\pm$0.70 \%  & 54753 & $\pm$1.28\% & 52083 & $\pm$1.21\%  & -1.33\% & -4.88\% & -6.14\%  \\ \hline
      8     & 53413 & $\pm$2.33 \%  & 53574 & $\pm$1.20\% & 51005 & $\pm$0.72\%  & -0.14\% & -4.51\% & -4.37\%  \\  \hline
    16      & 51302 & $\pm$1.13 \%  & 50841 & $\pm$0.99\% & 49107 & $\pm$1.44\%  & -0.78\% & -3.61\% & -4.36\%  \\  \hline
    32      & 47702 & $\pm$1.83 \%  & 47498 & $\pm$1.46\% & 45067 & $\pm$0.95\%  & -0.02\% & -5.39\% & -5.42\%  \\  \hline
    64      & 37049 & $\pm$0.83 \%  & 36687 & $\pm$1.16\% & 35861 & $\pm$1.00\%  & -0.83\% & -2.51\% & -3.31\%  \\  \hline
    128     & 25852 & $\pm$0.55 \%  & 25808 & $\pm$0.69\% & 25398 & $\pm$0.76\%  & -0.06\% & -1.54\% & -1.60\%  \\  \hline

    \multicolumn{10}{|c|}{2 workers} \\ \hline
    0    &132013  & $\pm$1.80\%  & 131029  & $\pm$1.97\% & 123723 & $\pm$0.48\%  & -0.68\% & -5.09\% & -5.73\%  \\\hline          
    1    &106254  & $\pm$0.61\%  & 105192  & $\pm$1.40\% & 99769  & $\pm$1.36\%  & -1.36\% & -4.95\% & -6.24\%  \\\hline          
    2    &104029  & $\pm$0.68\%  & 104104  & $\pm$1.35\% & 99469  & $\pm$0.39\%  & -0.18\% & -4.57\% & -4.74\%  \\\hline          
    4    &104776  & $\pm$1.21\%  & 105309  & $\pm$0.73\% & 99836  & $\pm$1.02\%  & +0.30\% & -5.37\% & -5.09\%  \\   \hline
    8    &101995  & $\pm$1.30\%  & 101960  & $\pm$0.56\% & 95991  & $\pm$0.78\%  & +0.24\% & -5.78\% & -5.55\%  \\   \hline
    16   &98139   & $\pm$0.44\%  & 96381   & $\pm$1.16\% & 92208  & $\pm$0.49\%  & -1.74\% & -4.32\% & -5.99\%  \\   \hline
    32   &89601   & $\pm$0.43\%  & 89253   & $\pm$0.62\% & 84653  & $\pm$1.51\%  & -0.16\% & -5.01\% & -5.17\%  \\   \hline
    64   &67910   & $\pm$1.31\%  & 67249   & $\pm$1.10\% & 64342  & $\pm$1.34\%  & -0.95\% & -4.40\% & -5.31\%  \\   \hline
    128  &46904   & $\pm$1.36\%  & 46821   & $\pm$0.49\% & 45525  & $\pm$0.33\%  & -0.12\% & -2.68\% & -2.80\%  \\   \hline
    \multicolumn{10}{|c|}{4 workers} \\\hline
    0    & 258531 & $\pm$3.90\%   & 257575 & $\pm$3.03\% & 243747 & $\pm$3.03\% & -0.37\% & -5.37\% & -5.72\%  \\\hline          
    1    & 203020 & $\pm$2.92\%   & 202384 & $\pm$2.45\% & 192917 & $\pm$2.45\% & -0.31\% & -4.68\% & -4.98\%  \\\hline          
    2    & 202407 & $\pm$2.61\%   & 202701 & $\pm$2.88\% & 190962 & $\pm$2.88\% & +0.15\% & -5.79\% & -5.65\%  \\\hline          
    4    & 203955 & $\pm$2.69\%   & 204389 & $\pm$3.70\% & 194757 & $\pm$3.70\% & +0.21\% & -4.71\% & -4.51\%  \\   \hline
    8    & 198404 & $\pm$3.43\%   & 197599 & $\pm$2.38\% & 187675 & $\pm$2.38\% & -0.41\% & -5.02\% & -5.41\%  \\   \hline
    16   & 188374 & $\pm$4.20\%   & 188913 & $\pm$2.48\% & 179768 & $\pm$2.48\% & +0.29\% & -4.84\% & -4.57\%  \\   \hline
    32   & 172736 & $\pm$2.38\%   & 169735 & $\pm$2.61\% & 161146 & $\pm$2.61\% & -1.74\% & -5.06\% & -6.71\%  \\   \hline
    64   & 135227 & $\pm$3.61\%   & 132880 & $\pm$1.76\% & 127736 & $\pm$1.76\% & -1.74\% & -3.87\% & -5.54\%  \\   \hline
    128  & 93900  & $\pm$2.62\%   & 93940  & $\pm$3.21\% & 91661  & $\pm$3.21\% & +0.04\% & -2.43\% & -2.38\%  \\   \hline
    \end{tabular}}
    \end{center}
\end{table}
\end{minipage}
\end{figure*}

\clearpage
\begin{minipage}{\textwidth}
\section{OpenSSL Example}\label{sec:openssl_example}
\begin{multicols*}{2}

\begin{changed}
The excerpts of code in \Cref{lst:OpenSSLExample} and
\Cref{lst:gcm_encrypt_user_data} show an example usage of \SDRoB with deeply
nested domains as explained in \Cref{sec:domainnesting}. The excerpts show a
simple file encryption server in an event-driven architecture that encrypts
client data using OpenSSL and stores the ciphertext on the server.

Upon receiving a client request the \texttt{event\_handler} function
(\Cref{lst:OpenSSLExample}) performs the following tasks:
\begin{enumerate}
    \item [\dCOne] Reads the encryption key from a file chosen by user and
      generates a random initialization vector (IV).
    \item [\dCTwo] Calls \texttt{gcm\_encrypt\_user\_data}
      (\Cref{lst:gcm_encrypt_user_data}), which:
    \item [\dCThree] Reads a plaintext message from the user via a file
      descriptor that corresponds to a communication socket.
    \item [\dCFour] Encrypts said plaintext with AES GCM using the OpenSSL's
      ``Envelope'' (EVP) API.
    \item [\dCFive] Finally (\Cref{lst:OpenSSLExample}), the
      \texttt{even\_handler} stores the ciphertext (and Galois/counter mode tag)
      for later retrieval.
\end{enumerate}

The objective to introducing rollback capability to the event handler is to
achieve the properties described in \Cref{sec:api}, namely to
\begin{inparaenum}
    \item protect the event handler running in the root domain from errors in
      \texttt{gcm\_encrypt\_user\_data}, e.g., the possible overflow to the
      plaintext buffer (\Cref{lst:gcm_encrypt_user_data}, \dNine),
    \item encapsulate the pointer to the OpenSSL context (\texttt{ctx}) within
      an outer, nested domain in such a way that OpenSSL's key objects remain
      inaccessible to \texttt{gcm\_encrypt\_user\_data} should it malfunction,
      and
    \item simplify error handling for the domain in which the OpenSSL code is
      run.
\end{inparaenum}
To this end, the event handler (running in the root domain) creates a persistent
domain for OpenSSL execution that is inaccessible from both the root domain and
any nested domain (\Cref{lst:OpenSSLExample}, \dOne).  This execution domain is
immediately deinitilized to invalidate its saved execution context to avoid
unintended rollbacks to the beginning of \texttt{event\_handler}. The actual
execution context for rollback (within another nested domain) is established
later.

The event handler (still running in the root domain) creates two additional data
domains (\Cref{lst:OpenSSLExample}, \dTwo and \dThree).  The first data domain
(\dTwo) is used to store the encryption key after it has been read from the
file, and the random initialization vector (\dCOne). Both the root domain and
the OpenSSL domain can access this domain area.

The second data domain (\dThree) is used for data that is shared between the
OpenSSL domain and the domain \texttt{gcm\_encrypt\_user\_data} will run in.
This corresponds to the third design option in \Cref{sec:api} with respect to
how the OpenSSL wrapper (\Cref{lst:openssl_wrapper}) is expected to operate.
The OpenSSL domain is granted access to the data domain as the data domain is
created (\dThree) The nested domain used to run the
\texttt{gcm\_encrypt\_user\_data()} function is first initialized, then granted
shared access to the data domain (\Cref{lst:OpenSSLExample}, \dFour).  The event
handler (still running in the root domain) uses this shared data domain to
allocate buffers that hold the plaintext, ciphertext, and the Galois/Counter
Mode (GCM) tag which are either used to communicate data to the nested domains,
or data back from them (\Cref{lst:OpenSSLExample}, \dFive).

The \texttt{gcm\_encrypt\_user\_data()} function is then invoked inside the nested domain (\Cref{lst:OpenSSLExample}, \dCTwo).
It then re-initializes the OpenSSL domain to set it to use the nested domain's saved execution context upon an abnormal domain exit (\Cref{lst:gcm_encrypt_user_data}, \dEight). This avoids the need to establish individual rollback points for each individual OpenSSL invocation that execute in the dedicated persistent domain. 

In case of an abnormal domain exit from \emph{either} domain the execution of the \texttt{gcm\_encrypt\_user\_data()} or OpenSSL is rolled back to the point of the second \texttt{sdrob\_init()} in \texttt{event\_handler} (\Cref{lst:OpenSSLExample}, \dFour).
The return value of that call (if other than \texttt{SUCCESSFUL\_RETURNED}) indicated the UDI of the domain that initiated the abnormal exit. When this occurs, the \SDRoB library has already destroyed the offending execution domain, but the caller uses the returned UDI to determine any remaining cleanup operations, such as destroying any remaining domains (\Cref{lst:OpenSSLExample}, \dSix and \dSeven).
Note that in \dSix, if the nested domain running \Cref{lst:gcm_encrypt_user_data} experienced and abnormal domain exit, the persistent OpenSSL domain still remains, but cannot be entered again until a new execution context for rollback is established. In principle the event handler could leave the domain intact (but deinitialized) at \dSix and reuse it the next time the event handler is called. For clarity, we show each domain explicitly destroyed in \Cref{lst:OpenSSLExample}. 
In \dSeven, the persistent OpenSSL domain experienced an abnormal exit. As it was initialized by \texttt{gcm\_encrypt\_user\_data()}  to use the calling domains execution context for rollback (\Cref{lst:gcm_encrypt_user_data}, \dEight) \SDRoB has automatically destroyed both the offending OpenSSL domain and the nested domain where \texttt{gcm\_encrypt\_user\_data()} executed.
Data domain are always left intact after rollback and must be explicitly destroyed.

On normal domain exit it is the responsibility of the event handler to deinitialize or destroy any domains before it exits, as explained in \Cref{sec:implementation_overview}. Similar to above, the event handler could choose to leave any of the domains intact (but deinitialize) to reuse them the next time it is entered.

\end{changed}
\end{multicols*}
\end{minipage}


\begin{figure*}
\noindent\begin{minipage}{\textwidth}
    \begin{lstlisting}[style=CStyle, label={lst:OpenSSLExample}, caption={An excerpt from an example file encryption server. The except shows the event handler code that has been augmented to perform the \SDRoB domain management, executing in the root domain. Some error handling has been omitted for brevity.}]
int event_handler(struct event_handler_args *args) 
{
    gcm_encrypt_user_data_args_t gcm_args;             // holds read-only arguments for gcm_encrypt_user_data()
    register gcm_encrypt_user_data_args_t *gcm_args_p  asm("r12") = &gcm_args;  // pointer passed across stacks
    register int ciphertext_len asm("r13");            // return value from nested domain held in callee-saved registers
    %*\tikzmark{onebegin}*)
    if(sdrob_init(OPENSSL_UDI, EXECUTION_DOMAIN | INACCESSIBLE_DOMAIN | RETURN_HERE) !=  SUCCESSFUL_RETURNED) {
        handleErrors(); /* If the OpenSSL domain initialization fails, don't continue*/
    }
    sdrob_deinit(OPENSSL_UDI);
    %*\tikzmark{oneend}*)
    %*\tikzmark{twobegin}*)
    if(sdrob_init(OPENSSL_PRIVATE_DATA_UDI, DATA_DOMAIN) !=  SUCCESSFUL_RETURNED) {
        sdrob_destroy(OPENSSL_UDI, NO_HEAP_MERGE);
        handleErrors();
    }
    sdrob_dprotect(OPENSSL_UDI, OPENSSL_PRIVATE_DATA_UDI, READ_ENABLE | WRITE_ENABLE);
    %*\tikzmark{twoend}*)

    %*\tikzmark{threebegin}*)
    if(sdrob_init(OPENSSL_SHARED_DATA_UDI, DATA_DOMAIN) != SUCCESSFUL_RETURNED) {
        sdrob_destroy(OPENSSL_PRIVATE_DATA_UDI, NO_HEAP_MERGE);
        sdrob_destroy(OPENSSL_UDI, NO_HEAP_MERGE);
        handleErrors();
    }
    sdrob_dprotect(OPENSSL_UDI, OPENSSL_SHARED_DATA_UDI, READ_ENABLE | WRITE_ENABLE);
    %*\tikzmark{threeend}*)
    %*\tikzmark{conebegin}*)
    gcm_args.key = sdrob_malloc(OPENSSL_PRIVATE_DATA_UDI, AES_GCM_KEY_LEN);
    if (read_key_from_file(key_p, AES_GCM_KEY_LEN, args->pathname) != 1){
        handleErrors();
    }
    
    gcm_args.iv = sdrob_malloc(OPENSSL_PRIVATE_DATA_UDI, AES_GCM_IV_LEN);
    if (RAND_bytes(gcm_args->iv, sizeof(AES_GCM_IV_LEN)) != 1){
        handlerErrors();
    }
    %*\tikzmark{coneend}*)
    %*\tikzmark{fourbegin}*)
    udi_t ret = sdrob_init(NESTED_DOMAIN_UDI, EXECUTION_DOMAIN| ACCESSIBLE_DOMAIN| RETURN_TO_CURRENT);

    if(ret ==  SUCCESSFUL_RETURNED) {
        sdrob_dprotect(NESTED_DOMAIN_UDI, OPENSSL_SHARED_DATA_UDI, READ_ENABLE | WRITE_ENABLE);
    %*\tikzmark{fourend}*)
        %*\tikzmark{fivebegin}*)
        gcm_args.ciphertext = sdrob_malloc(OPENSSL_SHARED_DATA_UDI,  CIPHER_TEXT_LEN);
        gcm_args.plaintext = sdrob_malloc(OPENSSL_SHARED_DATA_UDI, args->plaintext_len);
        gcm_args.tag = sdrob_malloc(OPENSSL_SHARED_DATA_UDI, TAG_SIZE);
        %*\tikzmark{fiveend}*)
        sdrob_enter(NESTED_DOMAIN_UDI);
        ciphertext_len = gcm_encrypt_user_data(gcm_args_p); %*\dCTwo*)
        sdrob_exit(); 
        sdrob_destroy(NESTED_DOMAIN_UDI, NO_HEAP_MERGE);
    } else {
        switch (ret) {
            %*\tikzmark{sixbegin}*)case NESTED_DOMAIN_UDI:
                sdrob_destroy(OPENSSL_UDI, NO_HEAP_MERGE); 
                sdrob_destroy(OPENSSL_SHARED_DATA_UDI, NO_HEAP_MERGE); 
                sdrob_destroy(OPENSSL_PRIVATE_DATA_UDI, NO_HEAP_MERGE);       
            %*\tikzmark{sixend}*)break;
            %*\tikzmark{sevenbegin}*)case OPENSSL_UDI:
                sdrob_destroy(OPENSSL_SHARED_DATA_UDI, NO_HEAP_MERGE); 
                sdrob_destroy(OPENSSL_PRIVATE_DATA_UDI, NO_HEAP_MERGE);
            %*\tikzmark{sevenend}*)break;
            default:
                abort();     
        }
        return OPERATION_FAILED;
    }

    // store ciphertext and tag for later retrieval  %*\dCFive*) 
    sdrob_destroy(OPENSSL_UDI, NO_HEAP_MERGE); 
    sdrob_destroy(NESTED_DOMAIN_UDI, NO_HEAP_MERGE);
    sdrob_destroy(OPENSSL_SHARED_DATA_UDI, NO_HEAP_MERGE); 
    sdrob_destroy(OPENSSL_PRIVATE_DATA_UDI, NO_HEAP_MERGE);
}
\end{lstlisting}
\begin{tikzpicture}[remember picture, overlay, thick]
    \draw[decorate,decoration={brace,amplitude=5pt,mirror}] ([shift={(-4pt,-2pt)}]pic cs:onebegin)
    -- ([shift={(-4pt,5pt)}]pic cs:oneend)
    coordinate[midway,xshift=-5pt](Btip);
    \draw[rounded corners] (Btip) -- +(0,0) node[left,none]
    {\scriptsize \dOne};

    \draw[decorate,decoration={brace,amplitude=5pt,mirror}] ([shift={(-4pt,-2pt)}]pic cs:twobegin)
    -- ([shift={(-4pt,5pt)}]pic cs:twoend)
    coordinate[midway,xshift=-5pt](Btip);
    \draw[rounded corners] (Btip) -- +(0,0) node[left,none]
    {\scriptsize \dTwo};

    \draw[decorate,decoration={brace,amplitude=5pt,mirror}] ([shift={(-4pt,-2pt)}]pic cs:threebegin)
    -- ([shift={(-4pt,5pt)}]pic cs:threeend)
    coordinate[midway,xshift=-5pt](Btip);
    \draw[rounded corners] (Btip) -- +(0,0) node[left,none]
    {\scriptsize \dThree};
    
    \draw[decorate,decoration={brace,amplitude=5pt,mirror}] ([shift={(-4pt,-2pt)}]pic cs:conebegin)
    -- ([shift={(-4pt,5pt)}]pic cs:coneend)
    coordinate[midway,xshift=-5pt](Btip);
    \draw[rounded corners] (Btip) -- +(0,0) node[left,none]
    {\scriptsize \dCOne};

    \draw[decorate,decoration={brace,amplitude=5pt,mirror}] ([shift={(-4pt,-2pt)}]pic cs:fourbegin)
    -- ([shift={(-4pt,5pt)}]pic cs:fourend)
    coordinate[midway,xshift=-5pt](Btip);
    \draw[rounded corners] (Btip) -- +(0,0) node[left,none]
    {\scriptsize \dFour};

    \draw[decorate,decoration={brace,amplitude=5pt,mirror}] ([shift={(-4pt,-2pt)}]pic cs:fivebegin)
    -- ([shift={(-4pt,5pt)}]pic cs:fiveend)
    coordinate[midway,xshift=-5pt](Btip);
    \draw[rounded corners] (Btip) -- +(0,0) node[left,none]
    {\scriptsize \dFive};

    \draw[decorate,decoration={brace,amplitude=5pt,mirror}] ([shift={(-4pt,-2pt)}]pic cs:sixbegin)
    -- ([shift={(-4pt,5pt)}]pic cs:sixend)
    coordinate[midway,xshift=-5pt](Btip);
    \draw[rounded corners] (Btip) -- +(0,0) node[left,none]
    {\scriptsize \dSix};

    \draw[decorate,decoration={brace,amplitude=5pt,mirror}] ([shift={(-4pt,-2pt)}]pic cs:sevenbegin)
    -- ([shift={(-4pt,5pt)}]pic cs:sevenend)
    coordinate[midway,xshift=-5pt](Btip);
    \draw[rounded corners] (Btip) -- +(0,0) node[left,none]
    {\scriptsize \dSeven};
\end{tikzpicture}
\end{minipage}
\end{figure*}

\begin{figure*}
\noindent\begin{minipage}{\textwidth}
    \begin{lstlisting}[style=CStyle, label={lst:gcm_encrypt_user_data}, caption={The \texttt{gcm\_encrypt\_user\_data} function reads a plaintext message from a descriptor provided as argument and encrypts the plaintext with AES GCM using OpenSSL's high-level, envelope (EVP) API. The call to the EVP API functions have been wrapped to execute in domain \texttt{OPENSSL\_UDI} as shown in \Cref{lst:openssl_wrapper} in \Cref{sec:api}.}]
/* The gcm_encrypt_user_data_args_t structure that is passed by reference 
   is actually allocated on the root domain stack and hence read-only. */
int gcm_encrypt_user_data(const gcm_encrypt_user_data_args_t *args) {
    ssize_t len = 0;
    ssize_t num_bytes_read = 0;
    ssize_t ciphertext_len = 0;
    EVP_CIPHER_CTX *ctx;

    if(sdrob_init(OPENSSL_UDI, EXECUTION_DOMAIN| INACCESSIBLE_DOMAIN| RETURN_TO_PARENT) !=  SUCCESSFUL_RETURNED) {   %*\dEight*)
        return GCM_ENCRYPT_FAILED;
    }

    if((ctx = EVP_CIPHER_CTX_new()) == NULL) {                             // create and initialize the OpenSSL context
        sdrob_deinit(OPENSSL_UDI);                                         // deinitialize domain before returning
        return GCM_ENCRYPT_FAILED;
    }

    if(EVP_EncryptInit_ex(ctx, EVP_aes_256_gcm(), NULL, NULL, NULL) != 1)  // initialize cipher for encryption
        goto err_out;

    if(EVP_CIPHER_CTX_ctrl(ctx, EVP_CTRL_GCM_SET_IVLEN, args->iv_len, NULL) != 1) //  set length of IV
        goto err_out;

    if(EVP_EncryptInit_ex(ctx, NULL, NULL, args->key, args->iv) != 1)      // load key and iv from private data domain
        goto err_out;
    %*\tikzmark{cthreebegin}*)
    while(num_bytes_read < plaintext_len) {  
        num_bytes_read += read(fd, plaintext, 1024); %*\dNine*)
    }
    %*\tikzmark{cthreeend}*)
    %*\tikzmark{cfourbegin}*)
    if(EVP_EncryptUpdate(ctx, NULL, &len, args->aad, args->aad_len) != 1)  // provide any additional authentication data
        goto err_out;

    if(EVP_EncryptUpdate(ctx, args->ciphertext, &len, args->plaintext, args->plaintext_len) != 1)
        goto err_out;
    ciphertext_len = len;

    if(EVP_EncryptFinal_ex(ctx, args->ciphertext + len, &len) != )         //  finalize the encryption
        goto err_out;
    ciphertext_len += len;

    if(EVP_CIPHER_CTX_ctrl(ctx, EVP_CTRL_GCM_GET_TAG, TAG_SIZE, args->tag) != 1)  // read the tag to the shared domain
        goto err_out;
    %*\tikzmark{cfourend}*)
    
    sdrob_deinit(OPENSSL_UDI);                                             // deinitialize domain before returning
    EVP_CIPHER_CTX_free(ctx);                                              // success, cleanup and return
    return ciphertext_len;

err_out:                                                                   // normal error occurred, cleanup and return
    sdrob_deinit(OPENSSL_UDI);                                             // deinitialize domain before returning
    EVP_CIPHER_CTX_free(ctx);
    return GCM_ENCRYPT_FAILED;
}
\end{lstlisting}
\begin{tikzpicture}[remember picture, overlay, thick]
    \draw[decorate,decoration={brace,amplitude=5pt,mirror}] ([shift={(-4pt,-2pt)}]pic cs:cthreebegin)
    -- ([shift={(-4pt,5pt)}]pic cs:cthreeend)
    coordinate[midway,xshift=-5pt](Btip);
    \draw[rounded corners] (Btip) -- +(0,0) node[left,none]
    {\scriptsize \dCThree};

    \draw[decorate,decoration={brace,amplitude=5pt,mirror}] ([shift={(-4pt,-2pt)}]pic cs:cfourbegin)
    -- ([shift={(-4pt,5pt)}]pic cs:cfourend)
    coordinate[midway,xshift=-5pt](Btip);
    \draw[rounded corners] (Btip) -- +(0,0) node[left,none]
    {\scriptsize \dCFour};
\end{tikzpicture}
\end{minipage}
\end{figure*}

\ifreviewer
\definecolor{darkraspberry}{rgb}{0.53, 0.15, 0.34}
\newcommand{\reviewercolor}[1]{{\color{darkraspberry} #1}}
\newcommand{\reviewer}[1]{\medskip\noindent\reviewercolor{#1}\par}

\clearpage

\setcounter{page}{1}


\noindent %
\textbf{Dear PC Members, dear Reviewers,} \\

Our submission "Unlimited Lives: Secure In-Process Rollback with Isolated
Domains" had previously been submitted to another conference, but was not
accepted for publication. Below we highlight how we extended and improved the
work over the previous submission. Original reviewer comments are highlighted
in \reviewercolor{color}. The full reviews are attached at the end. \\

\noindent %
\textbf{Improvement and Extension for USENIX Security 2023}

\reviewer{The paper proposes an API for compartmentalizing and isolating code and
data (or data only), so that rollbacks are performed relatively easy
later on.  Given that there is no automation involved, and the original
code needs to be manually altered, the proposed API needs to be evaluated
in terms of usability, effectiveness, and perhaps maintainability. What
makes the solution optimal (or good enough)? What experiments provide
evidence to support the above? What's the required effort to manually
compartmentalize applications with SDRoB? One case study is not even
remotely enough to draw conclusions. (Also, surprisingly, there is no
discussion re: the porting effort involved in the case of Memcached.)}

While the original submission only included the Memcached case study in
\Cref{sec:memcached}, this revised submission now provides two additional real-world use cases, NGINX in \Cref{sec:nginx} and OpenSSL in \Cref{sec:ssl}. Also, we show 
an additional, concrete example of using our API in \Cref{sec:openssl_example}. We point out how different applications would benefit from secure rollback in different
ways in~\Cref{sec:discussion}. Also, we describe different compartmentalization
schemes in the OpenSSL example of \Cref{lst:openssl_wrapper}.

We further evaluate the required effort to equip different applications with rollback capabilities 
and the pre-requisite isolation in \Cref{sec:memcached} and \Cref{sec:nginx}.
We account for the necessary code changes in the respective subsection in \Cref{sec:case_study}, summarized below: 
Our changes were limited to 2 source files in Memcached and 484 new lines of wrapper code.
In total, the changes amount to ~550 LoC of the 29K SLoC code base ($2\%$)
In NGINX, the required changes are limited to one file and 195 new lines of wrapper code.
In total, these changes amount to ~220 LoC of the 150K SLoC code base ($0.15\%$)
We developed a bespoke test application that makes use of rollback functionality with OpenSSL. 
The relevant portions of the application that deal with the domain management are shown in \Cref{sec:openssl_example}.

To the best of our knowledge, \SDRoB is the first work to provide roll-back
functionality for applications under attack on commercial off the shelf
processors (in difference to, e.g., CompartOS~\cite{Almatary2022}) and
based on readily available isolation mechanisms (cf.
\Cref{sec:background}). Related work on checkpointing and restoring
applications (e.g.,~\cite{Kashyap16, checkpointrestart, seccheck}, discussed in \Cref{sec:related_work}) follow
strictly weaker attacker models and are not directly comparable with our
work.

\reviewer{SDRoB seems to require from the developer to have a deep understanding
of the various parts of the code to be compartmentalized, interactions
between components, as well as what constitutes safety-critical state and
what not.  Or, in general, what needs to be isolated. Is this whole
approach tractable?  If yes, I'd like to see some qualitative or
quantitative evidence in support of the above. If the developer knows what
needs to be isolated, and also needs to write code for recovering from the
rollback, what's the whole purpose of SDRoB? Just to provide a wrapper for
spilling/filling register state and managing the PKU operations?}

In literature, the SFI mechanisms usually fall into two different categories: 
\begin{inparaenum}[1)]
  \item isolation of a shared library or kernel module; having a well-defined interface, these can be compartmentalized in a straight-forward manner, if not automatically,
  \item the solution provides an API that can be applied to compartmentalize real-world software, requiring manual effort and a good understanding of said software~\cite{Vahldiek-Oberwagner19, endokernel, Schrammel20}.
\end{inparaenum} 
Our proposal fits into the second category.  The reviewer's concern about "what
should be isolated within the application" is thus a general concern for SFI
solutions and not specific to our secure domain rollback mechanism. As we
discuss in \Cref{sec:domainlifecycle}, the isolation mechanism can be applied in
two ways: Protecting the application from a subroutine or protecting a
subroutine from its caller.  We provide a well-defined, flexible \SDRoB API,
that allows compartmentalizing applications in different ways.  We evaluate
the required effort to equip real-world applications with rollback capabilities
in \Cref{sec:case_study}. We also point out in~\Cref{sec:discussion} that
retrofitting applications written in memory-unsafe languages with rollback is a
cost-efficient way to improve an application's resilience to memory
vulnerabilities compared to a complete rewrite in a memory-safe
language. Potentially this also has less performance impact depending on which
memory safety languages are chosen. Furthermore, there is more to our
implementation than the reviewer suggests. In fact, SDRoB provides a
well-defined API for domain management, a reference monitor, rollback
capabilities, support for isolating global variables, split stack management,
and dedicated heap management, as well as the error detection and signal
handling mechanism as described \Cref{sec:implementation}.

\reviewer{There exist a wide range of solutions re: automated
compartmentalization.  For example, PtrSplit (CCS 2017), (uSCOPE, RAID
2021), and many, many more. The paper discusses none of these works and
provides no comparison between those approaches and the proposed one. What
makes SDRoB preferable? In which cases SDRoB provides increased
effectiveness?} 

We added a background section on SFI (\Cref{sec:sfi}) in the revised paper. 
It should be noted that none of the SFI solutions~\cite{Erlingsson06, Wahbe93, Castro09, Mao11, PtrSplit, Rivera16,Koning17,Hedayati19,Vahldiek-Oberwagner19,Melara19,Sung20,Lefeuvre21,Schrammel20,Wang20,Voulimeneas22,Kirth22,Jin22,Chen22, roessler2021} provide rollback capabilities as introduced by this work, they respond 
to detected domain violations by terminating the offending process. The novelty
of our idea is to provide secure rollback capabilities. Providing sufficient
in-process isolation is a \emph{prerequisite} for secure rollback, as explained
in \Cref{sec:threatmodel}, but not the main goal of this paper.

\reviewer{SDRoB relies on various additional hardening techniques that need to be
applied in the code that executes within an isolated domain, with CFI
being a critical/important one. However, the CFI requirement is not
discussed/analyzed properly. What kind of CFI scheme does SDRoB require?
Given that no CFI scheme is ideal, in terms of precision, and that
control-flow hijacking can still occur (within the isolated domain), what
prevents the attacker from mounting a code-reuse attack and remap parts of
the address space as they please, completely bypassing the isolation
offered by SDRoB? This is related to pt. 4 above. Even with data-only attacks, within
the isolated domain, one can tamper with arguments passed to system
calls, or, in general invoke system calls however they need. (This is
something within the scope of the threat model considered by the authors.)
What exactly prevents an attacker from duplicating a file descriptor in
Memcached and "steal" the I/O channel of another client? Similarly, what
prevents an attacker from closing an open file descriptor, or interacting
with the OS in a way that bypasses the confinement of SDRoB, or lead to a
DoS?} 

CFI plays two orthogonal roles in our approach. On one hand, as explained
in \Cref{sec:security_ev}, additional security measures are needed to ensure
that the main enforcement of \SDRoB isolation cannot be bypassed, including a
CFI scheme to prevent code reuse attacks against the reference monitor.  The
measures we discuss are not unique to the secure rollback mechanism in particular;
they are common to Intel PKU-based approaches, as we explain in~\Cref{sec:mpk} and ~\Cref{sec:discussion}, 
and would change if a different hardware mechanism for in-process isolation was used. On the other
hand, CFI mechanisms can be deployed by the developer in untrusted domains, to
support catching attacks and trigger the rollback, however the choice of such
CFI mechanisms is highly application-dependent and orthogonal to our design.

Similarly, the attacks described by the reviewer use system calls to break
domain isolation. These are common concerns for SFI mechanisms and can be
prevented by system call filtering, as discussed in \Cref{sec:security_ev}.
As we explained in \Cref{sec:highlevelidea} and \Cref{sec:security_ev}, 
the broader problem of preventing adversaries from tampering with OS resources is not a pre-requisite to facilitating in-process rollback, 
which is our main goal.

\reviewer{The performance evaluation is extremely limited. The paper characterizes
the impact of SDRoB just by the observed overhead on a single
application, under a specific compartmentalization scheme. The authors
should consider evaluating SDRoB on a wide range of applications, and also
take into consideration different compartmentalizations schemes (e.g.
coarser- vs. finer-grain ones). They should also characterize and analyze
how each part of SDRoB contributes to the overall overhead. For instance,
the 6.6/7.7 is mostly due to argument (and return value) copying?
Creating and destroying domains? PKU operations? CPU state spill and
refill? Something else?}

We now provide three different real-world use cases,
in \Cref{sec:memcached}, \Cref{sec:nginx}, and OpenSSL \Cref{sec:ssl}. of the
revised paper.  Also, we show an additional, concrete code example
in \Cref{sec:openssl_example}. Moreover, we described and evaluated different
compartmentalization schemes for the OpenSSL example
(\Cref{lst:openssl_wrapper}), highlighting the cost of copying arguments and
results between domains.  We profiled the cost of domain switching for NGINX,
and observed in \Cref{sec:nginx} that a significant portion of the domain switch
cost ($30\%-50\%$) comes from writing to the PKRU register due to pipeline
flushing.

\reviewer{According to the paper, the proposed system can recover the states when a
memory corruption (or an attack) is attacked in the isolated domain.
However, I think the authors do not clearly justify why this is necessary
for real applications. IMHO, the isolated modules are untrusted and the SFI
technique is to prevent the compromised modules from tampering with other
domains. In this case, why do we need to continue the execution of the main
applications? Even if the application can continue its execution, how to
ensure the functionality of the main application can not be affected? The
authors need to present promising application scenarios to motivate the
work.}

As we discussed in \Cref{sec:introduction} and \Cref{sec:discussion}, the secure
rollback of isolated domains is particularly suited for service-oriented
applications that require strong availability guarantees and may hold volatile
states such as client sessions, TLS connections or object caches. Redundancy and
load balancing can be used to minimize the impact of denial of service attacks,
but the loss of volatile state can still degrade service quality for clients,
which can be mitigated by our approach.  We presented real-world use cases to
support our idea: Memcached \Cref{sec:memcached} and NGINX \Cref{sec:nginx}. We
argue that if a malicious client request leads to memory corruption in NGINX,
the worker process may crash, and the master process restarts it; however, all
active connections of that worker are lost.  Also, the restart and loading time
for 10GiB of data into Memcached was about 2 minutes. Thus, an attacker who
successfully launches repeated attacks could knock out the Memcached service
without rollback. While this is clearly dominated by the loading time, even
applications without a such volatile state, but ultra-reliable low-latency
requirements can benefit from rollback.  We reproduced two real CVEs for NGINX
and Memcached to measure in-process rollback latency and compared it with
process restarting time and container restart time, respectively.
Finally, as outlined in \Cref{sec:domainlifecycle} and demonstrated in \Cref{lst:openssl_wrapper} as well as \Cref{sec:openssl_example}, our domain mechanism allows also to isolate library code that handles secrets from the rest of the program, with the ability to roll back parts of that exterior program state, still preserving integrity and confidentiality of the library data.

\newpage

\reviewer{Besides, the differences between this work and SFI are not clear. In the
security evaluations, the authors answered four research questions.
However, the last three questions are the ones solved by the SFI technique.
R2 and R3 are necessary requirements for an SFI system to implement the
isolation between domains. R4 is also needed to maintain the integrity of
the trusted entity of an SFI system (e.g., the trampoline which is
responsible for domain switching}

The four ``research questions'' are in fact requirements for the security and
effectiveness of our approach and \Cref{sec:security_ev} discusses how and why
these requirements are met by our solution.  We added a background section on
SFI (\Cref{sec:sfi}) in the revised paper.  The novelty of our idea is to
provide secure rollback capabilities. Providing sufficient in-process isolation
is a prerequisite for secure rollback as explained in \Cref{sec:threatmodel}.

\reviewer{
There are some common questions in the SFI system that are not addressed in this
paper. a) domain switch overhead b) the granularity of the data sharing between
domains }

As mentioned above, we extended our evaluation and profiled the domain switch overhead for NGINX in \Cref{sec:nginx} (a significant portion of the
domain switch cost ($30\%-50\%$) comes from writing to the PKRU register due to
pipeline flushing). However, it should be pointed out that our main contribution, secure rollback of isolated domains, is independent of the underlying SFI scheme and an implementation using a different SFI scheme would exhibit different domain switch overheads.

Concerning the granularity of data sharing, we discuss in \Cref{sec:domainlifecycle} and \Cref{sec:api} that there are different ways of sharing data between domains, depending on the compartmentalization strategy used: data may be explicitly copied to and from buffers allocated in the nested domain (at arbitrary granularity), or shared via a dedicated data domain (a separately managed heap memory, protected at page granularity). Smaller arguments and results may also be shared via registers and by default, data in the root domain is always readable by all other domains. All of these techniques are illustrated in \Cref{lst:sdrob_call}, \Cref{lst:openssl_wrapper}, and \Cref{sec:openssl_example}.

\bigskip
\noindent %
\textbf{Previous Reviews}
\tiny
\verbatiminput{20220605-reviews-ccs-anon.md}
\normalsize
\fi

\end{document}